\documentclass{article}

\usepackage{microtype}
\usepackage{graphicx}

\usepackage{algpseudocode}
\usepackage{algorithm2e}
\usepackage{caption}
\usepackage{subcaption}
\usepackage{booktabs} 
\usepackage{xspace}
\usepackage{xcolor}

\usepackage{hyperref}
\usepackage{makecell}

\usepackage{amssymb}
\usepackage{pifont}
\usepackage{comment}

\usepackage{color}

\definecolor{navy}{RGB}{0,0,128}

\definecolor{darkgreen}{rgb}{0.0, 0.5, 0.0}
\definecolor{amaranth}{rgb}{0.9, 0.17, 0.31}
\definecolor{azure}{rgb}{0.0, 0.5, 1.0}

\newcommand{\sref}[2]{\hyperref[#2]{#1 \ref*{#2}}}

\newcommand{\wsdnn}{WS-DNN\xspace}
\newcommand{\wsdnns}{WS-DNNs\xspace}
\newcommand{\fsreuse}{\textit{SubGraph}\xspace Reuse\xspace}
\newcommand{\fs}{\textit{SubGraph}\xspace}
\newcommand{\fsstationary}{\textit{SubGraph}\xspace Stationary\xspace}
\newcommand{\orcid}[1]{\href{https://orcid.org/#1}{\includegraphics[height=10pt]{orcid}}}

\definecolor{commt}{rgb}{0.2, 0.5, 0.2}

\newcommand{\secref}[1]{\S\ref{#1}}
\newcommand{\figref}[1]{Fig.~\ref{#1}}
\newcommand{\tabref}[1]{Tab.~\ref{#1}}

\newcommand{\insertFigure}[2]{
    \begin{figure}[t!]
        \centering
        \includegraphics[width=\linewidth]{Plot/#1.pdf}
	\vspace{-6mm}
        \caption{ #2}
	\vspace{-4mm}
        \label{fig:#1}
    \end{figure}
}

\newcommand{\insertFigureRatio}[3]{
    \begin{figure}[t!]
        \centering
        \includegraphics[width=#3\linewidth]{Plot/#1.pdf}
	\vspace{-3mm}
        \caption{ #2}
	\vspace{-2mm}
        \label{fig:#1}
    \end{figure}
}

\newcommand{\insertWideFigureRatio}[3]{
    \begin{figure*}[ht!]

        \centering
        \includegraphics[width=#3\textwidth]{Plot/#1.pdf}
	\vspace{-3mm}
        \caption{ #2}
	\vspace{-4mm}
        \label{fig:#1}
    \end{figure*}
}

\newcommand{\insertWideFigure}[2]{
    \begin{figure*}[ht!]

        \centering
        \includegraphics[width=\textwidth]{Plot/#1.pdf}
	\vspace{-8mm}
        \caption{ #2}
	\vspace{-6mm}
        \label{fig:#1}
    \end{figure*}
}

\newcommand{\squishlist}{
 \begin{list}{$\bullet$}
  { \setlength{\itemsep}{0pt}
     \setlength{\parsep}{0pt}
     \setlength{\topsep}{3pt}
     \setlength{\partopsep}{0pt}
     \setlength{\leftmargin}{1.5em}
     \setlength{\labelwidth}{1em}
     \setlength{\labelsep}{0.5em} } }

\newcommand{\squishnums}{
 \begin{list}{$\bullets$}
  { \setlength{\itemsep}{0pt}
     \setlength{\parsep}{3pt}
     \setlength{\topsep}{3pt}
     \setlength{\partopsep}{0pt}
     \setlength{\leftmargin}{1.5em}
     \setlength{\labelwidth}{1em}
     \setlength{\labelsep}{0.5em} } }

\newcommand{\squishlisttwo}{
 \begin{list}{$\bullet$}
  { \setlength{\itemsep}{0pt}
     \setlength{\parsep}{0pt}
    \setlength{\topsep}{0pt}
    \setlength{\partopsep}{0pt}
    \setlength{\leftmargin}{2em}
    \setlength{\labelwidth}{1.5em}
    \setlength{\labelsep}{0.5em} } }

\newcommand{\squishend}{
  \end{list}  }

\newcommand{\Accel}{{\textbf{\texttt{SushiAccel}}}\xspace}
\newcommand{\systemhw}{{\textbf{\texttt{SushiAccel}}}\xspace}
\newcommand{\systemsw}{{\textbf{\texttt{SushiSched}}}\xspace}
\newcommand{\systemAbs}
{{\textbf{\texttt{SushiAbs}}}\xspace}
\newcommand{\system}
{{\textbf{\texttt{SUSHI}}}\xspace}

\newcommand{\supernet}{{\textit{SuperNet}}\xspace}
\newcommand{\supernets}{{\textit{SuperNets}}\xspace}
\newcommand{\subnet}{{\textit{SubNet}}\xspace}
\newcommand{\subnets}{{\textit{SubNets}}\xspace}
\newcommand{\subgraph}{{\textit{SubGraph}}\xspace}
\newcommand{\subgraphs}{{\textit{SubGraphs}}\xspace}
\newcommand{\subgraphreuse}{{\textit{SubGraph}} Reuse\xspace}
\newcommand{\cmark}{\ding{51}}
\newcommand{\xmark}{\ding{55}}

\usepackage[accepted]{mlsys2023}

\mlsystitlerunning{SubGraph Stationary Hardware-Software Inference Co-design}

\begin{document}

\newif\ifcommenton

 \commentonfalse

\ifcommenton

\newif\ifshepherdon
\shepherdonfalse

\ifshepherdon
\newcommand{\shepherd}[1]{{\color{red}#1}}
\else
\newcommand{\shepherd}[1]{#1}
\fi

\twocolumn[
\mlsystitle{SubGraph Stationary Hardware-Software Inference Co-design}

\mlsyssetsymbol{equal}{*}

\begin{mlsysauthorlist}
\mlsysauthor{Payman Behnam}{equal,to}
\mlsysauthor{Jianming Tong}{equal,to}
\mlsysauthor{ Alind Khare}{to}
\mlsysauthor{Yangyu Chen}{to}
\mlsysauthor{Yue Pan}{to}
\mlsysauthor{Pranav Gadikar}{to}
\mlsysauthor{Abhimanyu Rajeshkumar Bambhaniya}{to}
\mlsysauthor{Tushar Krishna}{to}
\mlsysauthor{Alexey Tumanov}{to}

\end{mlsysauthorlist}

\mlsysaffiliation{to}{Georgia Institute of Technology, Atlanta, Georgia, USA}

\mlsyscorrespondingauthor{Payman Behnam, Jianming Tong}{payman.behnam@gatech.edu, jianming.tong@gatech.edu}

\mlsyskeywords{Machine Learning, MLSys}

\vskip 0.3in

\begin{abstract}
A growing number of applications depend on Machine Learning (ML) functionality and
benefits from both higher quality ML predictions and better timeliness (latency) at the same time.
A growing body of research in computer architecture, ML, and systems software literature
focuses on reaching better latency/accuracy tradeoffs for ML models.
Efforts include compression, quantization, pruning, early-exit models, mixed DNN precision, as well as
ML inference accelerator designs that minimize latency and energy, while preserving 
delivered accuracy. All of them, however, yield improvements for a single \shepherd{static} point in the latency/accuracy 
tradeoff space. We make a case for applications that operate in dynamically changing deployment scenarios,
where no single \shepherd{static} point is optimal. We draw on a recently proposed weight-shared \supernet mechanism 
to enable serving a stream of queries that uses (activates) different \subnets within this weight-shared construct.
This creates an opportunity to exploit the inherent temporal locality with our proposed \fsstationary (SGS) optimization. We take a hardware-software co-design approach with a real implementation of SGS in \systemhw{} and the implementation of a software scheduler \systemsw{} controlling which \subnets to serve and what to cache in real-time. 
Combined, they are vertically integrated into \system{}---an inference serving stack. 
{For the stream of queries \system yields up to 25\% improvement in latency, 0.98\% increase in served accuracy.
\system can achieve up to \shepherd{78.7\% off-chip energy savings}.}
\end{abstract}
]

\printAffiliationsAndNotice{\mlsysEqualContribution} 
\section{Introduction}
\label{sec:intro}

The number of applications leveraging Machine Learning (ML) functionality continues to grow, as ML is successfully applied beyond image classification \cite{image-classification}, object detection/recognition \cite{image_segmentation, object_recognition}, sentiment analysis \cite{sentiment_analysis}, and next word prediction \cite{next-word}. 
These applications are also increasingly latency sensitive.
Their interactive experience depends on what fraction of prediction tasks are satisfied within the application-specified latency budget (typically in the 10-100 ms interactive latency range). Examples of such applications include self-driving cars \cite{self_driving}, specifically the on-board software responsible for multi-modal sensory data processing, street sign detection \cite{street_sign}, pedestrian detection \cite{pedestrian_detection}, vehicle trajectory tracking \cite{vehicle_trajectory}, lane tracking \cite{lane_tracking}, and Intensive Care Unit stability score prediction~\cite{holmes}. 
These applications require the ability to serve trained ML models in a way that maximizes the fraction of queries completed within the application specified latency budget---defined as latency Service Level Objective (SLO) attainment. A unifying characteristic for this class of applications is that they simultaneously care about the quality (accuracy) and timeliness (latency) of ML inference served.

There has been a body of work successfully improving achievable latency/accuracy tradeoffs for specific Deep Learning models. Examples include multiple forms of quantization~\cite{bai2018proxquant,lq_net, apple_quantization,pwlq-eccv20}, mixed DNN precision~\cite{mixedprecdnn-mlsys21}, compression~\cite{squeezenet}, pruning~\cite{netpruning}, latency-aware neural architecture search~\cite{proxylessnas,lataware-nas}, just to name a few. However, fundamentally, all of these techniques optimize for a \shepherd{\textit{single static}} point in the latency/accuracy tradeoff space. Indeed, for a given deployment device, the outcome is typically a single static model that has a specific (latency, accuracy) tuple associated with it. We claim this is no longer sufficient.

We observe that the applications with acute latency/ accuracy sensitivity typically operate in \textit{dynamically} variable deployment conditions. These include variable query traffic patterns (e.g., variable number of patients triaged in the ICU or ER), on-device battery power level (e.g. bed-side compute or battery-powered edge device), and query complexity (e.g., autonomous vehicle (AV) navigation of sparse suburban vs dense urban terrain).

Under such variable deployment conditions, a choice of \textit{any} single static model from the latency/accuracy tradeoff space will be suboptimal. Indeed, a higher accuracy model may result in dropped queries during periods of transient overloads. The lower accuracy model may yield suboptimal prediction quality under low load---both unnecessarily under-performing. Inherently, the ideal solution would include dynamically picking a ``best-fit'' model from the latency/accuracy tradeoff space. 
For a specific latency constraint that varies over time, a just-in-time  choice of the highest accuracy model satisfying this constraint is preferred. 
Thus, the ability to switch (or navigate) between points in the latency/accuracy tradeoff space in real-time is intuitively required for such applications.

We identify one such mechanism that enables this --- weight-shared \supernets~\cite{ofa} (\secref{sec:back:ws}). 
This neural network construct consists of multiple convolutional neural networks (CNNs) sharing common model parameters. It simultaneously encapsulates ``deep and thin'' models as well as ``wide and shallow'' within the same structure without weight duplication. These \supernets can be used to activate different \subnets without explicitly extracting them into different independently stored models. This is highly efficient from the systems perspective, as it obviates the need to store these model variants separately (saving memory cost), and enables rapidly switching \subnets that are ``activated'' to serve different incoming queries.

On the hardware end, the need for real-time inference has led to a plethora of ML accelerators. A key optimization technique (e.g., ``dataflow"~\cite{chen2016eyeriss}) leveraged by most accelerators involves \textit{reusing} activations and/or weights across multiple computations, leading to architectures that can be classified as weight stationary, output stationary, input stationary, row stationary, and hybrid variations of these~\cite{chen2016eyeriss}.
These dataflows rely on neural network layers, specifically 2D convolutions, to be compute-bound.
One challenge of serving \subnets with diverse shapes, however, as we identify, is the memory-bound nature of some of the \subnets (smaller FLOPS/Byte).

To address this challenge, we make a key observation that the weight-shared \supernet mechanism inherently results in queries activating commonly shared \textit{\subgraph}s within 
the same \supernets structure\footnote{We define \subgraph as a subgraph consisting of any subset of weights from the \supernets connected together into a graph}.
Furthermore, we note a significant amount of \textit{temporal locality} in the weights of the \supernets re-used \textit{across} queries. We identify this as an opportunity for a new kind of data reuse, which we name \texttt{\fsstationary (SGS)}  optimization---a technique we haven't seen used or proposed by any existing accelerator. We realize the benefits of SGS by implementing hardware caching support for weight reuse at the granularity of neural network \subgraphs.

In addition to SGS-aware hardware implementation, we co-design an SGS-aware query scheduler that decides (a) which \subnets to activate for each query and (b) which \subgraphs to cache. We propose an algorithmic approach to make these control decisions based on (a) a query's specified accuracy constraint and 
(b) the current state of the accelerator (which we abstract). 
We demonstrate that these control decisions benefit from hardware state awareness, as baseline state-unaware caching leaves room for improvement.
Finally, we propose an abstraction that enables the query scheduling policy to generalize, while remaining accelerator state-aware. 
The abstraction is captured by a black-box table (\figref{fig:system_overview_new}) that exposes the latency of activating a \subnet $i$ as a function of a currently cached \subgraph $j$. 
We instantiate the concept of \fsstationary (SGS) cross-query optimization in our vertically integrated inference serving stack, \system{}, which includes 
(a) \systemhw{}---a real FPGA implementation of hardware support for SGS-aware weight-shared \supernet inference, and
(b) \systemsw{} to make real-time control decisions on a stream of queries executed on \systemhw{}, sequentially deciding for each query \subnet $i$ to activate and (periodically) \subgraph $j$ to cache on the accelerator.

\systemhw{} and \systemsw{} combined in \system enable \textit{agile} navigation of the latency/accuracy tradeoff space, reaching better latency/accuracy tradeoffs by leveraging the key property of ``cross query'' temporal locality inherent to weight-shared \supernets with what we believe to be the first hardware-software co-design for weight-shared inference.

The key contributions of this paper can be summarized as follows:
\squishlist
    \item a concept of \subgraph Stationary (SGS) approach for hardware acceleration of DNN inference on weight-shared \supernets.
    \item \systemhw{}---a real SGS-aware FPGA implementation, with a simulator and design space exploration tools.
    \item \systemsw{}---a software query scheduler that operates in SGS-aware fashion, controlling which \subnets to activate and \subgraphs to cache in real time.
    \item \system{}---a hardware-software co-designed inference serving stack, vertically integrating \systemhw{} and \systemsw{}.
    \item \shepherd{\systemAbs{}}---an abstraction that generalizes SGS-aware query scheduling to arbitrary accelerators, while retaining implicit accelerator state awareness.
\squishend
{Combined, \system is able to achieve up to 25\% query serving latency improvement with 0.98\% accuracy improvement. \system can also save a significant amount of off-chip energy (78.7\%) in simulation with realistic board configurations. }

\section{Background and Motivation}
\label{sec:back}
We start with a background on weight-shared neural networks in \secref{sec:back:ws}. Then we motivate and expose the opportunity for hardware support of weight-shared supernet inference (\secref{sec:back:hw}). The need for hardware-software co-design follows from challenges in \secref{sec:back:challenge}. The hardware-software abstraction in \secref{sec:back:sw} is introduced for generality.

\subsection{Weight-Shared Deep Neural Networks (\wsdnns)}
\label{sec:back:ws}
\begin{figure}[t]
    \centering
    \subfloat[\small \subnets \& \subgraphs concepts. \label{fig:subnetsupernet}]{{\includegraphics[width=0.53\linewidth]{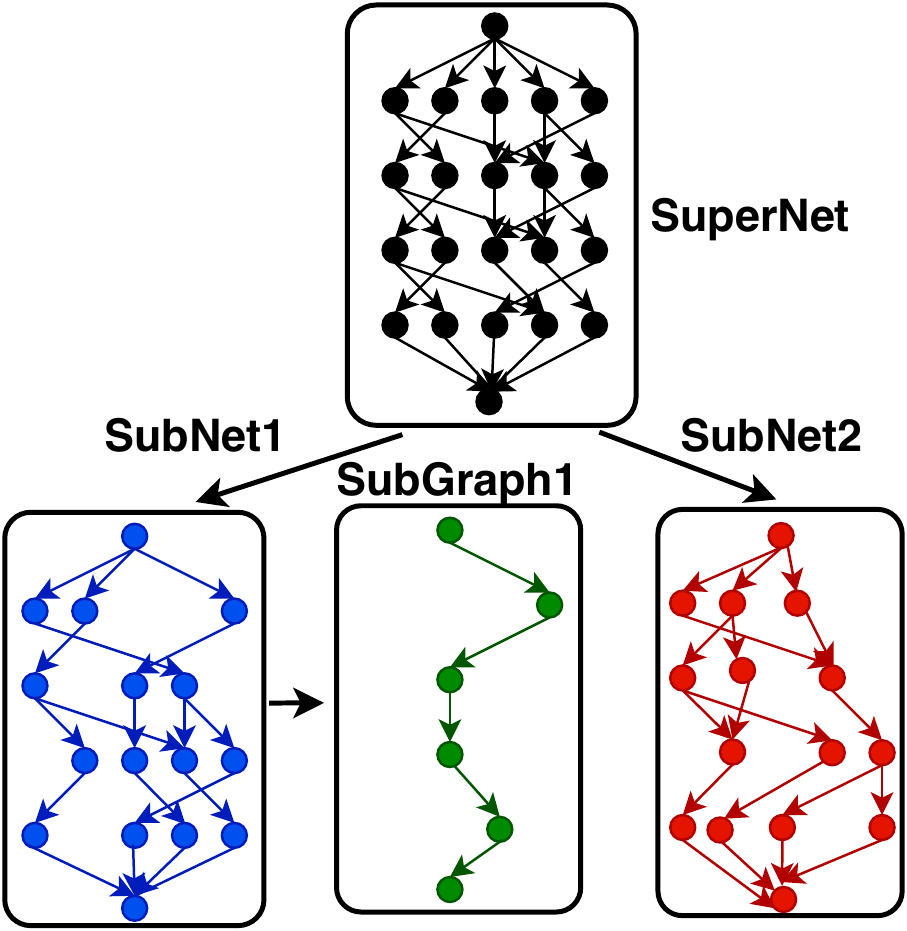}} }
    \hfill
    \subfloat[\small 
         Latency-Acc. tradeoff \label{fig:LatAccTrade}]{{\includegraphics[width=0.45\linewidth]{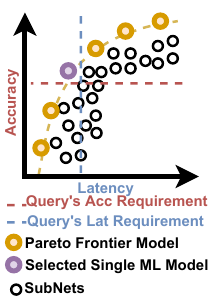}}} \hspace{-2mm}
    \vspace{-5mm}
    \caption{{\wsdnn properties.}}
    \vspace{-5ex}
\end{figure}
Recent advances in deep learning propose weight-shared deep neural networks \cite{ofa,compofa,bignas} that propose \supernet structures can be used to enable inference on Deep Neural Networks (DNNs) across a diverse set of deployment scenarios (both dynamic and static). Weight-shared DNNs (\wsdnn) induce a rich trade-off between accuracy and latency \shepherd{(Fig.~\ref{fig:LatAccTrade})}. The inference in \wsdnn fundamentally changes the traditional view of optimizing inference latency, which was focused on a single forward pass query. Instead, \wsdnn's inference \shepherd{makes it possible} to satisfy the latency-accuracy requirements for a \textit{stream of queries} with each query potentially requesting a different point in the trade-off space. This positions \wsdnns as a salient candidate for a variety of applications \cite{lat_acc_app1, lat_acc_app2, lat_acc_app3} and inference-serving systems \cite{infaas} that benefit from navigating latency/accuracy trade-off. The key property of these networks is that different DNNs (\subnet), which may differ in several elastic dimensions, including depth and width, partially share their weights as part of a single large DNN (\supernet). As a result, the \supernet contains all other \subnets within it \shepherd{ (Fig.~\ref{fig:subnetsupernet})}. These \subnets can be directly used to render predictions without any further re-training. To get predictions from a specific \subnet, elastic dimensions are specified in order to select appropriate weights from the \supernet for the forward pass.

These elastic dimensions typically include specification of the depth, the number of filters/channels of each convolutional layer and kernels. The elastic dimensions of the neural net architecture of the \supernet are exploited to attain elasticity. A typical \supernet architecture such as OFAResNet, OFAMobileNet is organized as a collection of stages. Each stage consists of repeating blocks, such as a Bottleneck block in OFAResNets. Each block in turn contains multiple convolution layers. The depth elastic dimension selects top $k \in [2;4]$ blocks per-stage of the \supernet. The expand ratio (another elastic dimension) selects top $k$ kernels of the convolution layer in each block. As a result, the smallest \subnet's weights are shared by all other \subnets and the weights of the largest \subnet contain all other \subnets within it. Hence, there's always some amount of common weight sharing between \subnets, with cardinality of overlap ranging from the smallest to the largest \subnet.

\subsection{Need for Hardware Support for \wsdnn Inference}
\label{sec:back:hw}

The goal of hardware acceleration for ML inference is to serve a query with minimal latency and maximal accuracy. 
This goal becomes even more pronounced for \wsdnn inference, where each query may be served with different latency/accuracy requirements \shepherd{(Fig.~\ref{fig:LatAccTrade})~\cite{ofa, compofa}}.

Achieving this goal is challenging due to memory-boundedness of some of the convolutional layers~\cite{kao2022flat, siu2018memory}. This is especially true for the more recent smaller models that have lower arithmetic intensity (FLOPS/Byte) and when they are deployed on bandwidth-constrained embedded boards~\cite{wang2019benchmarking,wei2019overcoming,chen2016eyeriss, jokic2020improving,siu2018memory,chen2022communication}.

\insertFigureRatio{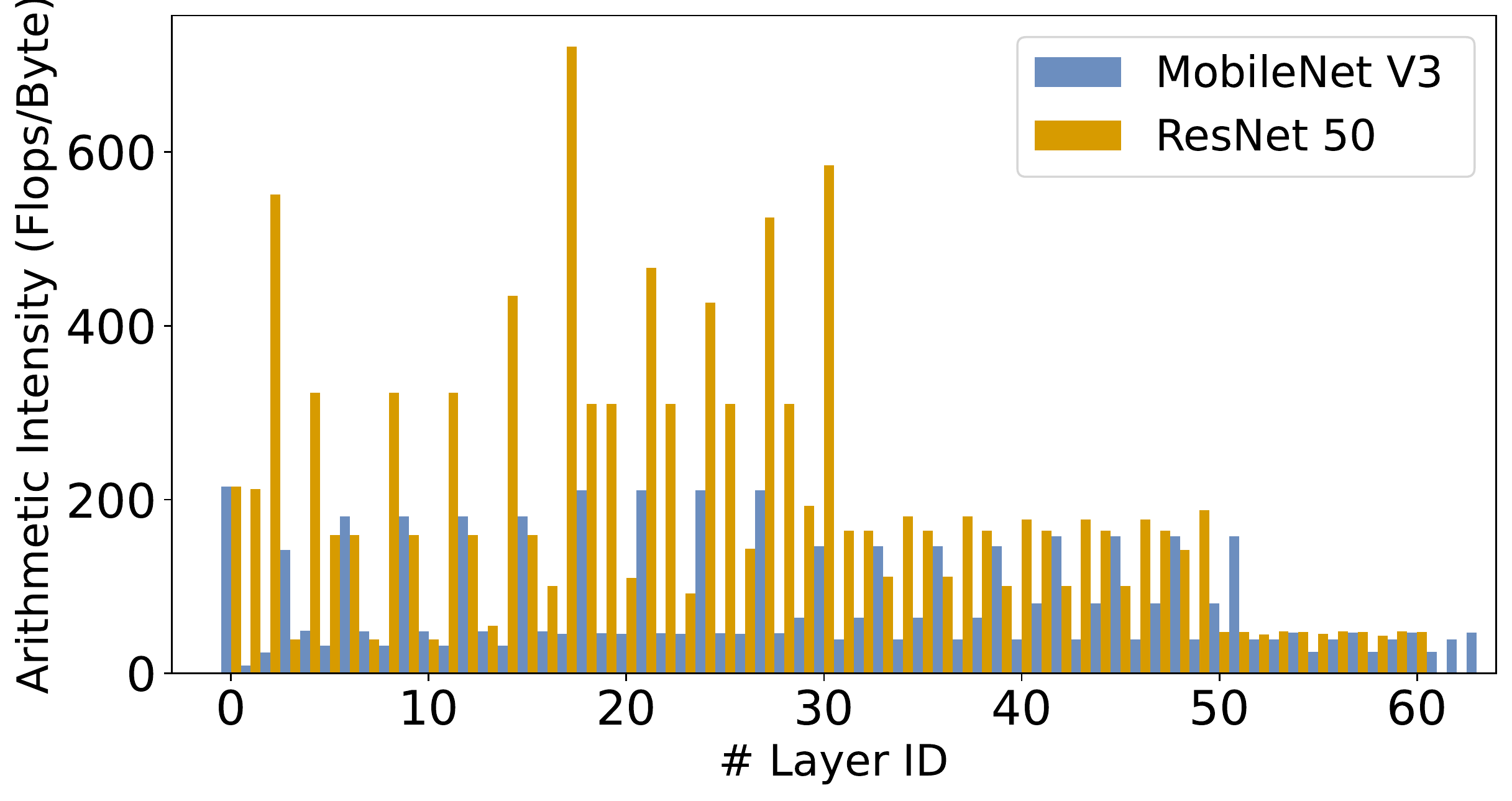}{Arithmetic intensity for different layers of various DNNs.
Lower arithmetic intensity leads to relatively higher \textit{memory} intensity in MBV3 and ResNet50's latter layers.\vspace{-3 ex}}{0.9}

We quantify this in 
Fig.~\ref{fig:motivation}, where we observe that a large fraction of convolution layers running on a canonical edge accelerator are memory-bound~\footnote{\shepherd{In the same network,} relatively lower arithmetic intensity corresponds to higher chances of becoming memory bound.}.
 
This is problematic, since a significant portion of end-to-end inference latency and energy consumption comes from memory-bound layers, given the high latency and energy cost of data movement from memory to the on-chip storage~\cite{chen2016eyeriss, yuan2021forms}. 

Hence, \shepherd{for the same amount of FLOPS} it is very important to convert memory-bound layers to compute-bound in order to reduce end-to-end inference latency and energy consumption.

To do so, we leverage our key insight that \wsdnn inference on a stream of queries exhibits temporal locality. As different queries use different \subnets, many of them reuse the same weights shared among those \subnets, by design. We employ this insight to help convert memory-bound layers to be more compute-bound. 
Conceptually, this can be accomplished by reusing the shared weights used by previous queries for the next query in a stream, knowing that they all activate \subnets within the same shared \supernet structure. This creates an opportunity for reuse \textit{across queries}, in sharp contrast to techniques commonly explored and exploited in the computer architecture community for a \textit{single} query for intra-model optimizations, such as weight-stationary, row-stationary, input-stationary, and output-stationary~\cite{chen2017using,chen2016eyeriss, fleischer2018scalable, venkatesan2019magnet}. 

We call this novel reuse as \fsreuse, as common shared weights form a \subgraph (e.g., created as the intersection of computational graphs of any two served \subnets). Note that in this paper we distinguish between \subgraphs and \subnets. \shepherd{\subnet is a subset of a \supernet that can be used for forward-pass inference to serve a query, while a \subgraph is a subset of \subnet. Note that any \subnet is a \subgraph, but not vice versa.}

A natural way to leverage \fsreuse is to have a dedicated cache in the hardware accelerator. However, it comes with several challenges that we discuss in \secref{sec:back:challenge}.

\vspace{-2ex}
\subsection{Design Challenges in \wsdnn Inference Specialized Hardware}
\label{sec:back:challenge}
The proposed specialized hardware for \wsdnn-inference exploits the temporal locality and enables \fsreuse. 
However, assigning a dedicated on-chip buffer comes with both software and hardware challenges.

\textbf{Hardware Challenges:} 
\shepherd{Due to the resource-restricted nature of many deployment devices,
the cache size may be too small to cache entire \subnets. Thus, the hardware must operate at a \textit{finer caching granularity} of arbitrary \subgraphs instead.
Deciding the size of the dedicated on-chip buffer is non-trivial.} 

Small buffer size leads to marginalizing the ability to exploit temporal locality. 
Larger dedicated on-chip buffer limits compute area as well as other on-chip buffer sizes that are leveraged for weight/row/input/output stationary optimizations.

\shepherd{Furthermore, the \fsstationary depends on the compute/memory boundness of the convolution workload, which is further related to the off-chip bandwidth and throughput of the hardware. Therefore, the variation of the bandwidth and throughput will also affect the best cache size, which introduces more factors for consideration in the trade-off space.}

\textbf{Software Challenges:} 
We argue that the latency of served \subnets depends on the \subgraph cached in the on-chip buffer. 
Fig.~\ref{fig:motiv_software_sched} provides a toy example to illustrate that: 
(a) a deep and thin \subnet gets a lower latency with a cached \subgraph containing more layers compared to other cached \subgraphs with fewer layers and wider bottleneck blocks, and 
(b) a wide and shallow \subnet achieves lower latency with a cached \subgraph with wider and fewer layers (matching its shape). 
This creates two challenges in software:
(a) \subnet selection decision to serve the current query must be \textit{aware} of the currently cached \subgraph (state), and 
(b) the cached \subgraph itself should be updated based on previously served \subnets for optimized latency. 
In other words, the software needs to make \textit{cache-state aware} decisions to select the appropriate \subnet and update the cached state based on temporally local (e.g., recent) \subnets  that were used to serve recent queries.

\vspace{-2ex}
\subsection{Hardware-Agnostic Software Scheduling}
\label{sec:back:sw}
One final goal is to achieve generalizability for the software scheduler while retaining accelerator state awareness. 
The scheduler policy design could then generalize to any hardware that is able to support \wsdnn inference. 
Hence, there is a need to decouple the scheduler from the hardware, i.e., the change in the hardware should not require any changes in the scheduler policy code. 
We propose an abstraction between the software scheduler and the hardware accelerator that exposes latencies of serving a set of \subnets over a set of cached \subgraphs. We show that this gives the policy sufficient information about the hardware state in an accelerator-agnostic fashion. We discuss the mechanism of achieving this while managing the spatial complexity of such a lookup table in \secref{sec:design}.
We instantiate this mechanism in \systemsw{}, which we can now develop and improve upon independently on any hardware accelerator.

\begin{figure}[t!]
 \vspace{-2mm}
    \centering
\includegraphics[width=0.9\columnwidth]{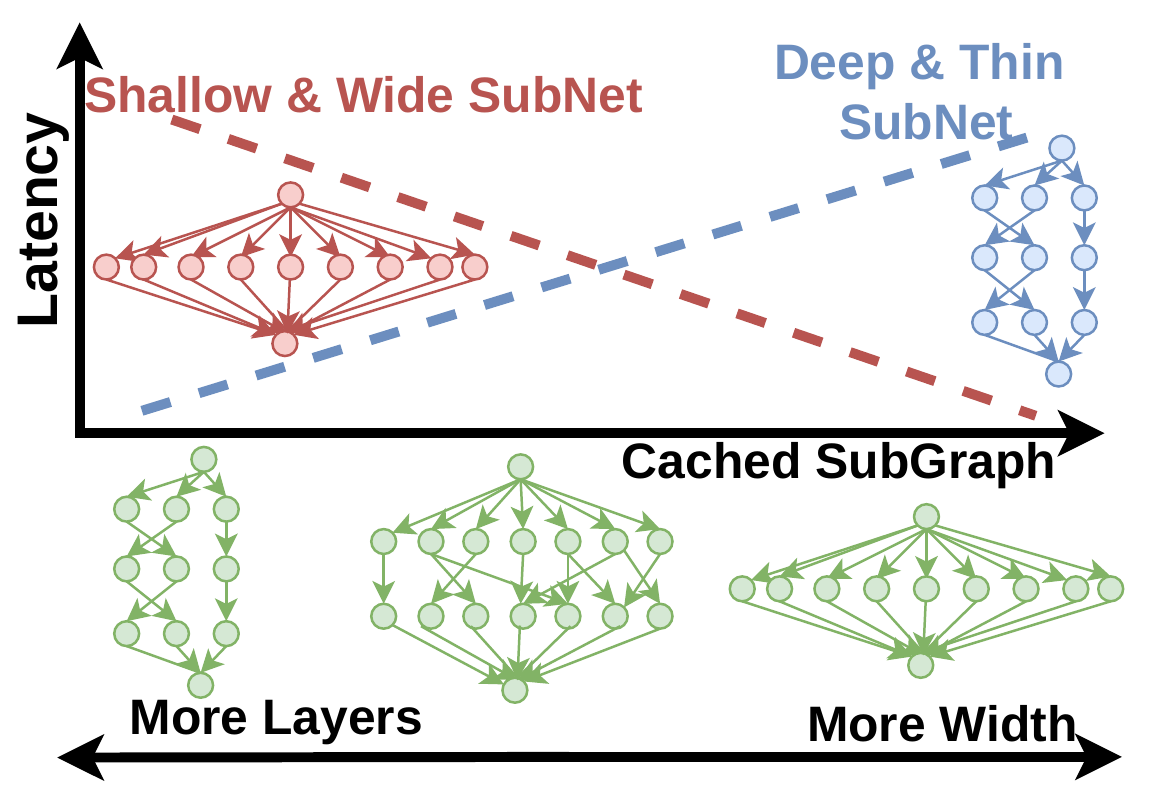}
 \vspace{-3mm}
    \caption{\small 
         Latency of two different \subnets as a function of different cached \subgraphs. 
         Different cached \subgraphs are optimal for different served \subnets with a non-trivial relationship based on the similarity of NN architecture parameters. \systemsw captures this similarity with a distance measure in \secref{sec:design}. 
         } \label{fig:motiv_software_sched}
\end{figure}

\section{System Design \& Architecture}
\label{sec:design}
\insertWideFigure{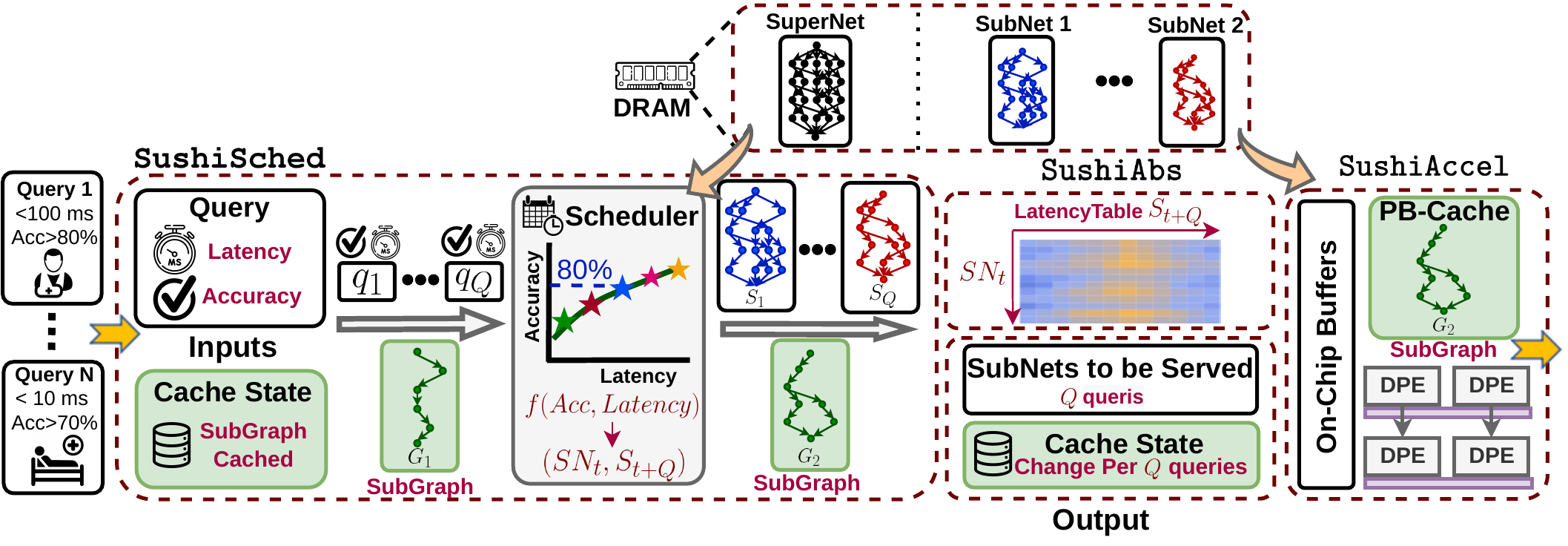}{System architecture overview. Given a stream of queries annotated with (Accuracy, Latency) pairs  $q_1,..,q_Q$ and the current cache state $C_1$, the scheduler chooses the \subnet to be served $SN_i$ for each $i$`th query and next cache state $G_2$ after every $Q$ queries.}

\system serves a stream of queries with different latency/accuracy requirements. 
It consists of three major components --- scheduler (\systemsw), abstraction (\systemAbs), and accelerator (\systemhw) as shown in Fig.~\ref{fig:system_overview_new}. \system exploits novel \fsreuse enabled via the interaction of its three components to serve queries with higher accuracy subjected to latency constraints or lower latency subjected to accuracy constraints. 

We describe our proposed \systemAbs and \systemsw below. \systemhw is described in \secref{sec:acc_Imp} in detail. The terminology used in this paper is captured in Fig.~\ref{fig:terminology}.
\insertFigure{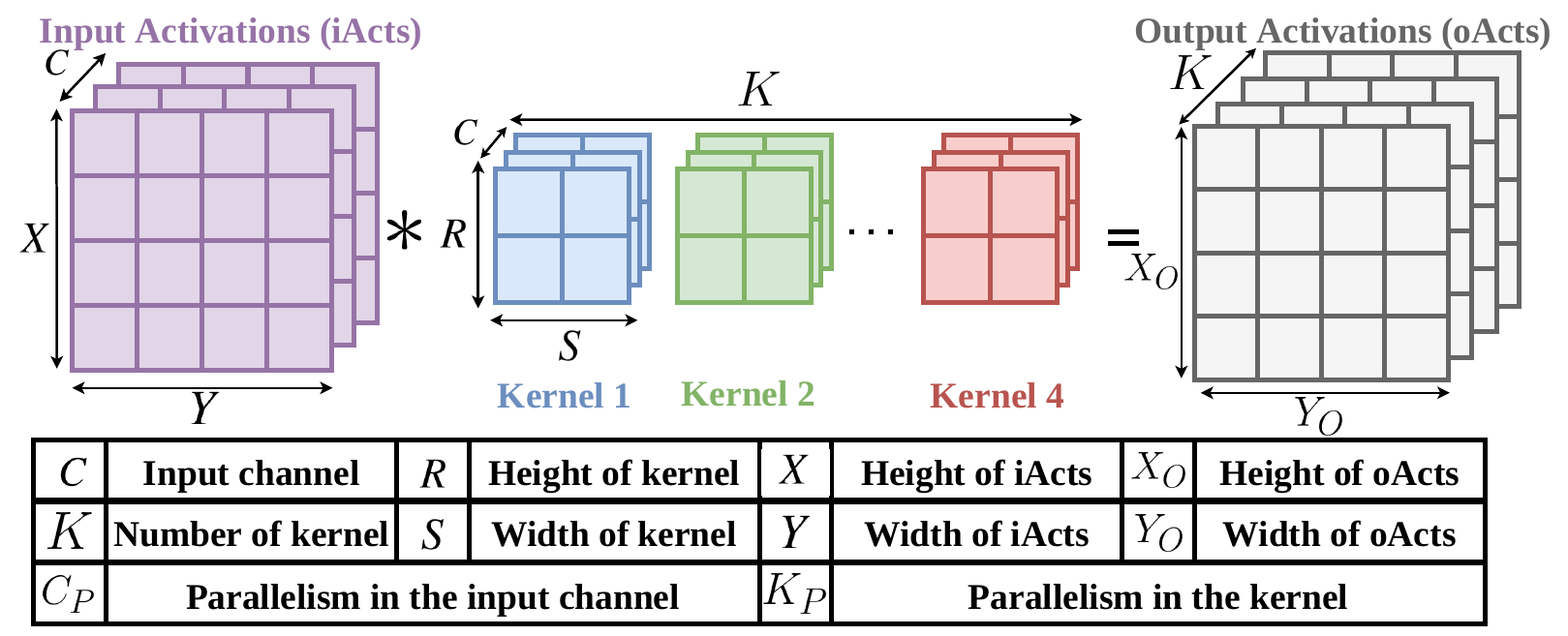}{\system terminology and variable definitions.\vspace{-2ex}}

\vspace{-2ex}
\subsection{\system's System Architecture}
We describe the interaction between \system's components. 
\figref{fig:system_overview_new} demonstrates a query path in \system. The query enters the system with a certain latency and accuracy constraint. Then, the \systemsw makes a two-part control decision. 
First, it selects an appropriate \subnet (i.e., $SN_t$) that can serve the current query $q_t$. It makes this subnet selection with the help of \systemAbs.
\systemAbs provides the scheduler with the ability to perform latency estimation when a specific \subnet is served with a given \subgraph cached. \systemAbs exposes this state in an accelerator-agnostic fashion. 
 
Second, \systemsw decides the next cached-\subgraph. The exact algorithm for this control decision is described in Alg.~\ref{alg:scheduler-algorithm}.

\systemsw control decision is then enacted by \systemhw. The selected \subnet, next cached-\subgraph, and query-data are sent to the \systemhw. \systemhw performs inference of the query using the selected \subnet. Model weights that are not already SGS-cached as part of the cached \subgraph are fetched from off-chip to on-chip buffer space. Finally, the accelerator returns the results of performing inference on \subnet to \systemsw and enacts the \subgraph caching control decision.

\vspace{-3ex}
 \subsection{Abstraction}
 \vspace{-2ex}
\systemAbs abstracts the ability to perform latency estimation for a given \subnet as a function of a cached \subgraph in an accelerator-agnostic fashion. It enables \systemsw to make cached\subgraph aware control decisions. As these control decisions are performed on the critical path of the query, this enabling abstraction must be efficient both w.r.t. space (R1) and time (R2).

Indeed, the set of all possible cached-\subgraphs is exponentially large for \wsdnns ($>> 10^{19}$) ~\cite{ofa}. 
Thus, to achieve (R1), the abstraction limits the set of all possible cached \subgraphs to a significantly smaller  set $\mathcal{S}$, such that $|\mathcal{S}| << 10^{19}$. 
The size of \subgraphs in $\mathcal{S}$ are selected to be close to the cache size.

Hence, at any point in time, \systemhw always caches \subgraphs from  $\mathcal{S}$ and \systemsw also selects a \subgraph to cache from $\mathcal{S}$ as well. 
The abstraction achieves (R2) by using a lookup table data structure with \subnets as rows and \subgraphs as columns.
{Hence, it takes the least amount of time to get latency-estimate of \subnet $i$ for a given \subgraph $j$.} The size of the lookup table is given by $O(|\mathcal{S}|.|\mathcal{X}|) \approx O(|\mathcal{S}|)$ where $\mathcal{X}$ denotes the set of serving \subnets, since we expect $O(|\mathcal{X}|) \approx O(1)$.

 \vspace{-2ex}
\subsection{\systemsw Design}

 \vspace{-4ex}
\RestyleAlgo{ruled}
\begin{algorithm}
\caption{Scheduling Algorithm}\label{alg:scheduler-algorithm}
\textbf{Input:}{ \subnet to be served $SN_i$, $i$ $\in$ $[1...N]$, \subgraph to be cached $G_j$, $j$ $\varepsilon$ $[1...M]$, Latency table $L[i][j]$.}

\KwResult{\subnet to be served and \subgraph to be cached.}
\textbf{Calculate} \subnet to be served for every query 

$q_t = (A_t, L_t)$, $t$ $\in$ $[0...Q]$ and \subgraph to be cached every $Q$ iterations;

$AvgNet$ = [0,0,0...0];
$CacheState = \varnothing$;

\While{$q_t$}{
  \eIf{policy == STRICT$\_$ACCURACY}{
    $id_x$ = argmin$_{latency}$($L[i][CacheState]$
     
     $\forall$$i$ $\in$ $[0...N]$ s.t. $SN_i$.accuracy $>=$ $A_t$)\;
  }{
      $id_x$ = argmax$_{accuracy}$($L[i][CacheState]$
      
      $\forall$$i$ $\in$ $[0...N]$ s.t. $SN_i$.latency $<=$ $L_t$)\;
    }
    \For{every $Q$ queries}{
    $AvgNet$.update($SN_{id_x}$, $Q$) 
    
    CacheState = argmin$_{Dist}$(Dist(G$_{j}$,AvgNet)) 
   
    $\forall$$j$ $\in$ $[0...M]$)\;}
  }
\end{algorithm}

On the software side, the scheduler receives a stream of queries, where each query is annotated with an (Accuracy, Latency) pair, denoted $(A_t, L_t)$. 
In this section we will describe exactly how the scheduler makes its \subnet selection and \subgraph caching control decisions.

{\bf Per-query \subnet ($SN_{t}$) Selection.}
As shown in Fig.\ref{fig:system_overview_new}, the scheduler decision is guided by two primary considerations: 

($i$) serve strictly higher accuracy and 
($ii$) serve strictly smaller latency, which can be specified by the user. In case of strictly higher accuracy, the scheduler can choose from the feasibility set of all \subnets with accuracy $\geq A_t$.
\system serves a \subnet that has minimum latency among all the \subnets that have accuracy $\geq$ $A_t$. Note that, it may be possible that the served latency might not satisfy the latency constraint of $\leq$ $L_t$. 
In case of strictly lesser latency, the scheduler serves a \subnet that has maximum accuracy among all the \subnets that have latency $\leq$ $L_t$. Similarly, it is possible that the served accuracy might not satisfy the accuracy constraint of $\geq$ $A_t$. Notice that the accuracy for a given \subnet is fixed, whereas the latency depends on the \subgraph cached into the PB. The scheduler employs a $Latency-Table$ to get the latency values for \subnet given a cache state.

{\bf Across-query \subgraph Caching ($S_{t+Q}$).}
The scheduler needs to decide what \subgraph to cache after every $Q$ queries ($S_{t+Q}$). To make this decision, the scheduler needs to represent the \subgraphs and \subnets, use the information from the past $Q$ queries, and predicts the next \subgraph that should be cached into the PB.

{\bf Encoding \subgraph NN Architecture.}
The scheduler represents both the \subnets and the \subgraphs as a vector as shown in Fig.~\ref{fig:subnetcache}. The scheduler uses the number of kernels $K_i$ and the number of channels $C_i$ of every layer $i$ to create a vector of size $2N$ for $N$ layered neural network. For instance, the vectorized representation for a 3-layered neural network would be $[K_1, C_1, K_2, C_2, K_3, C_3]$.

{\bf Amortizing Caching Choices.}
The scheduler keeps a running average of the past $Q$ \subnets that were served by the scheduler as shown in Fig.\ref{fig:subnetcache} (middle). The running average serves as a good indicator of the kernels and the channels that were frequently used in the \subnets that were served for the past $Q$ queries. If some kernels or channels were frequently used in the past $Q$ \subnets, the values corresponding to these kernels or channels will be high in the vectorized representation. Notice that, the running average can be considered as an approximation of the intersection operation, but with more information. Doing intersection purely loses the information for the kernels and the channels that were frequent but not present in all the \subnets; however, averaging helps us to preserve this information.

{\bf Predicting the Next \subgraph ($S_{t+Q}$).}
The scheduler employs the distance from the running average of the past $Q$ queries to predict the next \subgraph to be cached as shown in Fig.\ref{fig:subnetcache}. The scheduler caches the \subgraph that has the minimum distance from the average \subnet. Minimum distance ensures that the most frequent kernels and channels will be cached into the PB. In case fitting all of them is not possible, minimum distance from average \subnet ensures that we are picking the best fit \subgraph in terms of frequently occurring channels and kernels in the \subnets served by the scheduler.
The algorithm for performing both the scheduler decisions is described briefly in Algorithm~\ref{alg:scheduler-algorithm}. \systemsw receives input from the user including $\subgraphs, \subnets, LatencyTable$. AvgNet is the running average of the served \subnets. The cache state is set to a random \subgraph initially. The \systemsw decides the \subnet to be served for a given query when the accuracy is a hard constraint i.e. serving strictly better accuracy. The \systemsw can also decide the \subnet to be served if the latency is a hard constraint i.e. serving strictly lesser latency. It updates the running average of the \subnets. Finally, the \systemsw determines the \subgraph that is closest to the AvgNet and caches {\color{blue}{it}} into the PB.

\begin{figure}[tb]
\begin{center}
\includegraphics[scale=0.44]{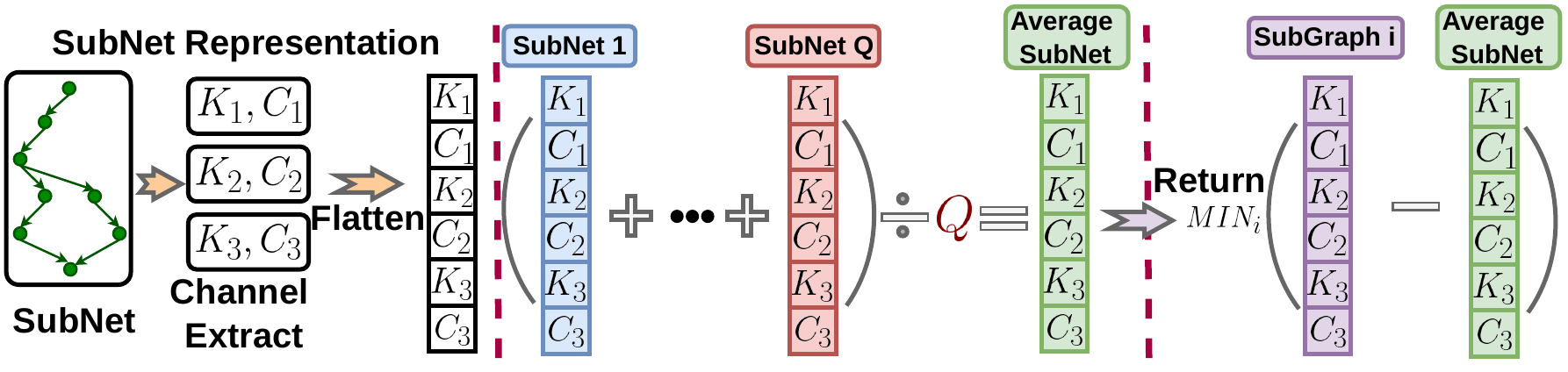}
    \caption{ The scheduler represents each neural network as a vector using the number of kernels and channels for each layer. The scheduler maintains a running average of the \subnets that were served for the past $Q$ queries. For every $Q$ queries, the scheduler caches the \subgraph that is the closest to the average \subnet.}
 \label{fig:subnetcache}
\end{center}
\vspace{-6mm}
\end{figure}
\section{\systemhw Implementation}
\label{sec:acc_Imp}
\label{sec:hw}

\subsection{Hardware Design Challenges}
\label{sec:hw_challenge}
As discussed earlier in \secref{sec:back} and \secref{sec:design}, to support \fsstationary, we propose to augment DNN accelerators with a custom cache called Persistent Buffer (PB).
{The introduction of PB leads to a new design space because it} competes for a finite on-chip buffer capacity (that needs to be partitioned across input activation, weight, and output activation tiles, and also shared weights).

To guarantee the best performance of hardware design on such {a design space}, we have to develop the parameterizable hardware template with the support of different hardware configurations.

\insertWideFigureRatio{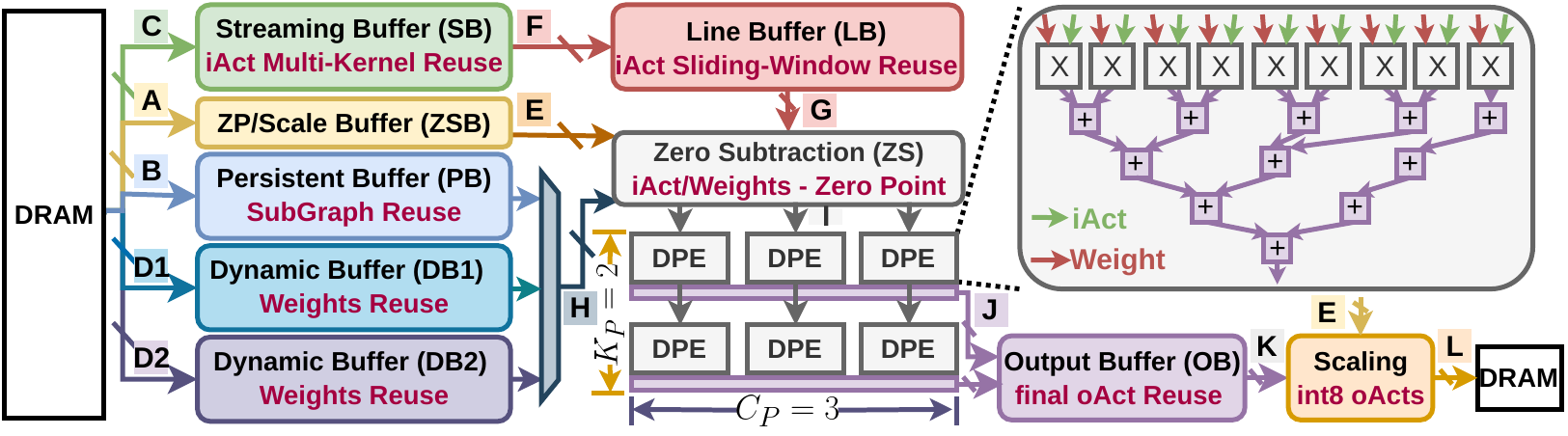}{The overall \Accel architecture ($K_P=2, C_P=3$)}{0.9}

\vspace{-3ex}
\subsection{Architectural Components}
In this part, we introduce components of \systemhw (\figref{fig:FastSwitch_architecture_shrink}) and how it supports all proposed data reuse in \figref{fig:workload_data_reuse}.

\begin{figure*}[t]
    \centering
    \vspace{-3mm}
    \subfloat[iAct Sliding Window Reuse. \label{fig:sliding_win_reuse}]{{\includegraphics[width=0.56\columnwidth]{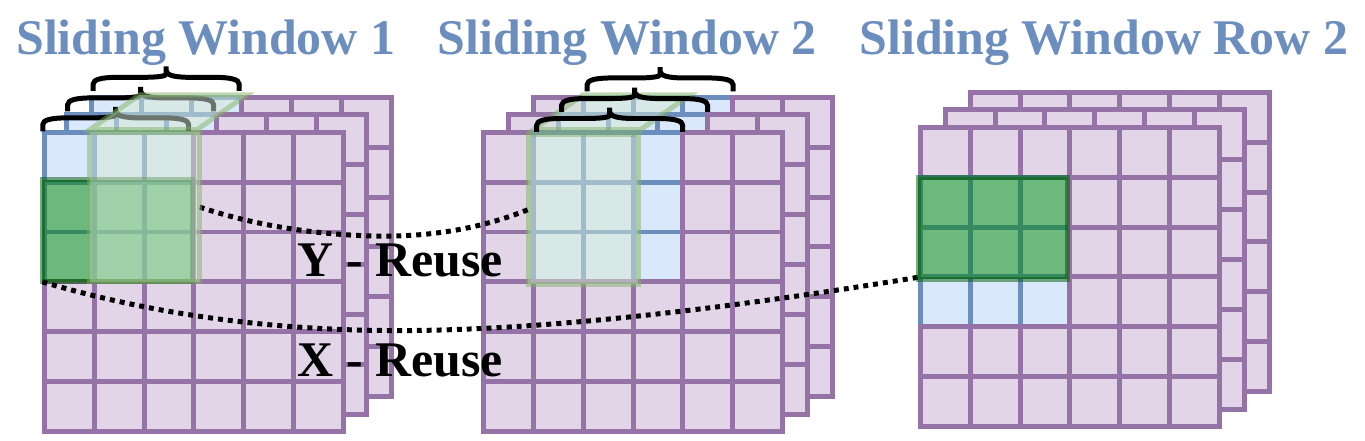}}} 
    \subfloat[iAct Multi-Filter Reuse. \label{fig:kernel_reuse}]{{\includegraphics[width=0.5\columnwidth]{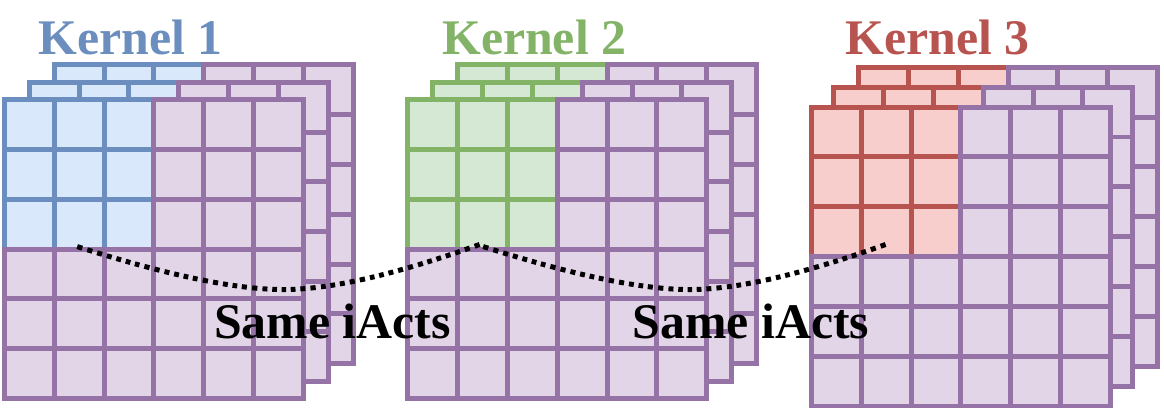}}}
    \subfloat[oAct Reuse.\label{fig:Partial_Sum_Reuse}]{{\includegraphics[width=0.54\columnwidth]{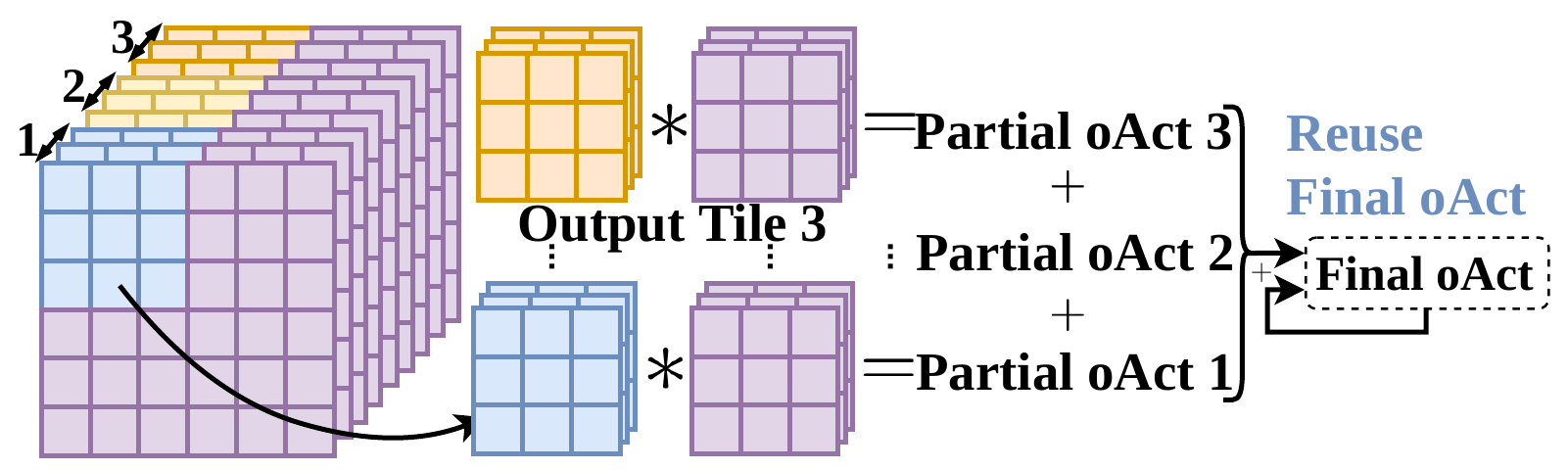}}} 
    \subfloat[\fsreuse \label{fig:subnet_reuse}]{{\includegraphics[width=0.48\columnwidth]{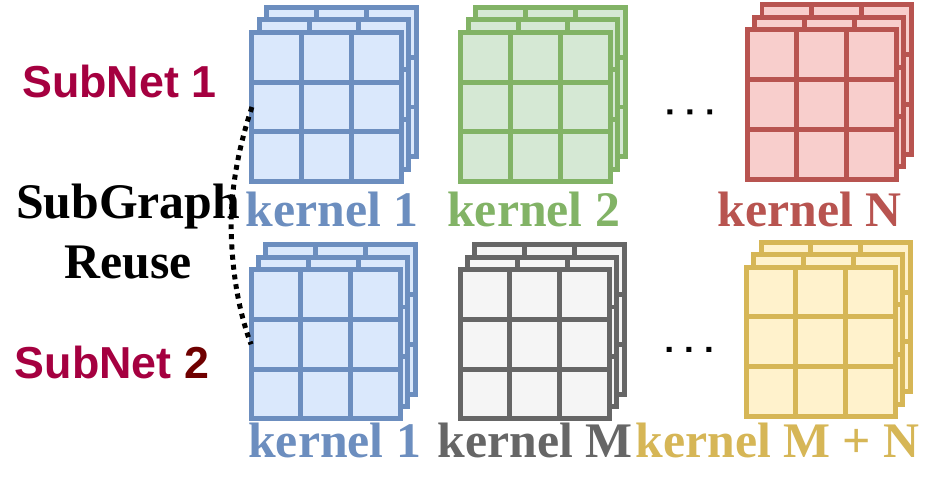}}}
    \vspace{-2mm}
    \caption{Data reuse opportunities in serving different \subgraphs leveraged within \Accel.}
    \label{fig:workload_data_reuse}
\end{figure*}

\subsubsection{Compute Array}
\label{sec:compute}
\quad \textbf{Dot Product Engine (DPE).} 
The key building block of DNN accelerators is the ability to compute 
\textit{dot-products}. 
For example, the Google TPU systolic array~\cite{jouppi2017datacenter} computes fixed-size dot products in each column by keeping weights stationary and forwarding (streaming) inputs from one column to the other,
NVDLA~\cite{nvdla} employs dedicated dot product engines (DPEs) of size 64, while flexible accelerators~\cite{kwon2018maeri,sigma} have DPEs of configurable sizes (enabled via all-to-all connectivity between the buffers and PEs).
In this work, we picked fixed-size DPEs of size 9. Larger kernels will be breakdown into a serial of $3\times3$ kernels and get flattened across the multipliers for reduction using the adder tree. As for small kernels ($1\times1$), $C$ dimension will be flattened across multipliers to leverage input channel parallelism.

\textbf{Parallelism.}
To further increase the throughput, we instantiate a 2D array of DPEs to boost the throughput by leveraging parallelism and reuse as shown in the \figref{fig:workload_data_reuse}. As for the parallelism, the number of row indicates the total number of kernels being processed in parallel in DPE Array, i.e. kernel-level parallelism ($K_P$). While the number of column stands for total number of input activation (iAct) channels being processed in parallel, i.e. channel-level parallelism ($C_P$). 

Both iActs and weights take the same interface to save the wire cost and improve scalability. In the vertical axis, both weights and iActs pass through DPEs of different rows in the store-and-forward fashion. During the weights forwarding, DPE will keep targeted weights stationary. Then, iActs will be streamed and get processed. In the horizontal axis, we replicate the same DPE independently to process different iActs channels and add an extra adder tree to reduce results from DPEs in the same row.

\subsubsection{On-chip Buffers and Supported Data Reuse}
We designed a custom on-chip buffer hierarchy to both store data in the layout preferred by the DPE array and support reuse opportunities not leveraged by the DPE array. The entire on-chip storage is divided into multiple separate buffers for different types of data as illustrated by different colors in \figref{fig:FastSwitch_architecture_shrink}.

\textbf{Persistent Buffer (PB)}.
The PB is designed to enable \fsreuse. 
For example, \Accel loads the \subgraph (kernel 1) in \figref{fig:subnet_reuse} from off-chip memory only once and stores it inside PB, such that it could be reused when switching between \subnet 1 and \subnet 2. 

\textbf{Dynamic Buffer (DB)}.
The DB is a typical on-chip storage to store the distinct weights of the requested \subnet. By adopting a PB, only non-common weights need to be fetched from the off-chip to the on-chip storage. For example, in \figref{fig:subnet_reuse}, all kernels except the common part (kernel $2$ to kernel $N$) will be loaded into DB when targeting at \subnet 1, and will be replaced by kernel $M$ to kernel $M+N$ when switching into \subnet 2. The DB is implemented as a ping-pong buffer, as indicated by DB1 and DB2 in \figref{fig:FastSwitch_architecture_shrink}, to hide the latency of fetching distinct weights from the off-chip DRAM.

\textbf{Streaming Buffer (SB)}. SB is designed to store entire iActs and support \textit{iAct Reuse - Multiple kernels.} (\figref{fig:kernel_reuse}).

\textbf{Line Buffer (LB)}. LB works as a serial to parallel conversion~\cite{wang2021ac} because the line buffer takes a single pixel from SB and moves it internally. Therefore, iActs data among different sliding windows will be reused inside the LB, i.e. LB supports \textit{iAct Reuse - Sliding Window Overlap} (\figref{fig:sliding_win_reuse}). We augment the naive line buffer to support stride by enabling sliding windows skipping.  

\textbf{Output Buffer (OB)}. OB provides in-place accumulation for oActs of different channels such that only the final oActs will be sent off-chip to save data movement of partial sums.

\textbf{ZP/Scale Buffer (ZSB)}. ZSB serves as the on-chip storage for zero point and scale for quantized inference. 
\begin{figure}[t]
    \centering
    \vspace{-3mm}
     \subfloat[Multi-query timeline. \label{fig:SR_Bonus}]{{\includegraphics[width=0.495\columnwidth]{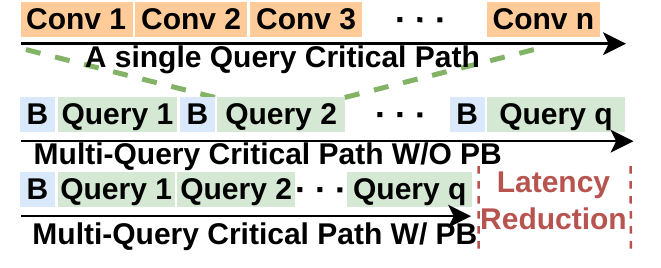}}}
    \vspace{-2mm}
    \subfloat[Single Conv timeline. \label{fig:dataflow_sgl_layer}]{{\includegraphics[width=0.485\columnwidth]{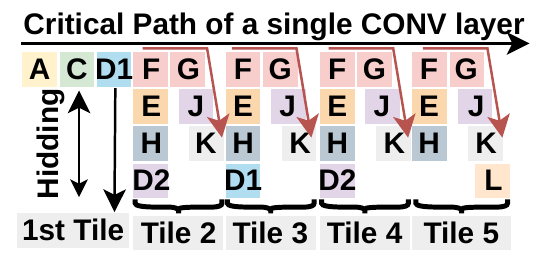}}} 
    \caption{\Accel dataflow overview.} 
    \label{fig:overll_dataflow}
    \vspace{-8.5mm}
\end{figure}
\subsection{\Accel Dataflow}

\subsubsection{Latency Reduction from Inter-Query Dataflow}
The inter-query processing timeline of \Accel is shown in \figref{fig:SR_Bonus} where stage B indicates the movement of the common \fs from off-chip to on-chip PB. 
The latency saving of \Accel comes from eliminating the redundant off-chip \fs access, as illustrated in \figref{fig:SR_Bonus} where \Accel reduces common \fs off-chip access (stage B) to only once in the critical path instead of multiple times in design w/o PB. 

\subsubsection{Hiding Latency from Intra-layer Dataflow}
Within each convolution layer, \Accel processes a convolution layer in the granularity of weight tiles shown in \figref{fig:dataflow_sgl_layer}. Different stages (i.e., A-L) are defined in \figref{fig:FastSwitch_architecture_shrink} that represent the movement of specific data.
To further hide off-chip data access latency from critical path, we implement a double distinct weights buffer (ping-pong dynamic buffers DB1 and DB2 shown in \figref{fig:FastSwitch_architecture_shrink}) to hide the off-chip latency of fetching distinct weights behind the computation latency.  This is indicated by stages D1 and D2 that are hidden from stages F-G-J-K shown with arrows in \figref{fig:dataflow_sgl_layer}.

\section{Experimental Results}
\label{sec:exp}

\subsection{System Setup}
\label{sec:exp:setup}

\noindent \textbf{Workload:} We choose weight shared version of ResNet50 and MobV3 as two \supernets~\cite{ofa}. To evaluate \system with full range on the pareto-frontier, we pick a sequence of 6 and 7 \subnets from ResNet50 and MobV3, respectively. 

The sizes of ResNet50 \subnets range from the [7.58 MB, 27.47 MB] while the sizes of MobV3 \subnets range from [2.97 MB, 4.74 MB]. Shared weights take up 7.55 MB and 2.90 MB for ResNet50 and MobV3, separately~\footnote{Weights, input activations, and zero points are quantized to int8, and the quantization scale is quantized into int32.}. \subnets are obtained using the procedure mentioned in OFA~\cite{ofa}.

\noindent \textbf{Metrics:} Latency in this section refers to the end-to-end serving latency of a given model, while accuracy refers to the top-1 accuracy.
Both accuracy and latency are defined for \subnets only. \subgraphs are only used for the caching purpose as a subset of \subnets. 

\noindent \textbf{Architecture Analytic Model:}
We have developed an analytic model which estimates the behavior of \systemhw to explore design space by configuring the architecture with parameters. 

Our model accurately predicts the latency trend of \systemhw using profiled latency of \systemhw on both workloads, enabling us to perform an exhaustive search of all parameter combinations within specified constraints.
This approach allows for the identification of optimal configurations for improved performance in both simulation and real-world deployment.

\textbf{Roofline Analysis} \noindent We also extended a roofline analysis tool to study the effect of PB on the boundness of \systemhw under different workloads.

\noindent\textbf{Deployment Platforms:}
We implemented the proposed \Accel on two FPGA including ZCU104 (5 W) and Alveo U50 (75 W).  
We compare our \systemhw w/ PB and w/o PB with Xilinx DPU and CPU (Intel i7 10750H, 45 W). 

\noindent \textbf{Scheduler Simulator:}
We have developed \systemsw, which runs on the CPU and guides the \systemhw on how to serve the current query and (a) what \subgraph to serve and (b) \subnet to be placed in PB. 

\vspace{-1ex}
\subsection{\system Impact on Arithmetic Intensity}

\begin{figure}[t]
    \centering
    \vspace{-0.15cm}
   \subfloat[ResNet50. \label{fig:latency_accuracy_resnet50}]{{\includegraphics[width=0.52\linewidth]{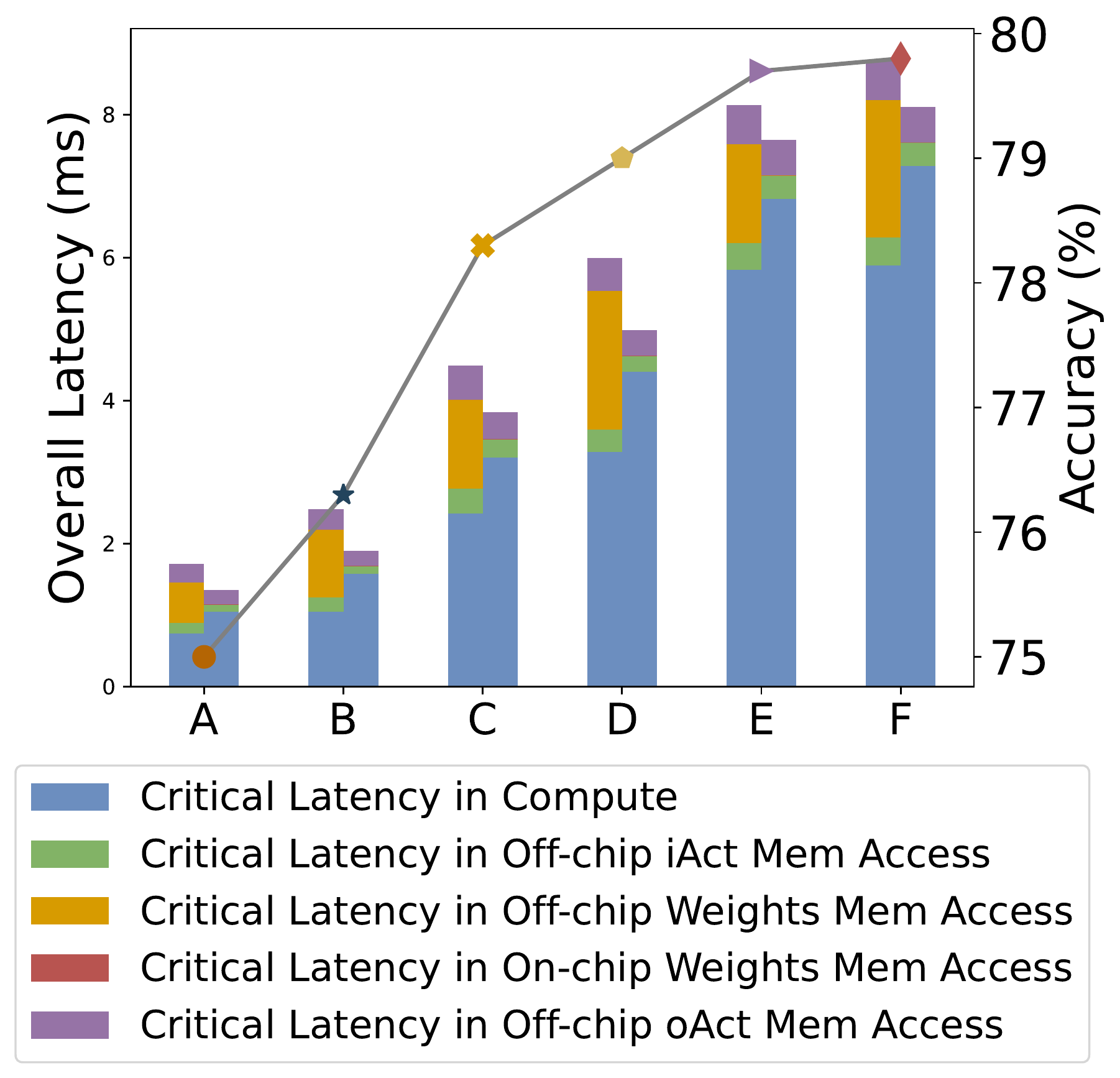}}}
    \subfloat[MobV3. \label{fig:latency_accuracy_mbv3}]{{\includegraphics[width=0.53\linewidth]{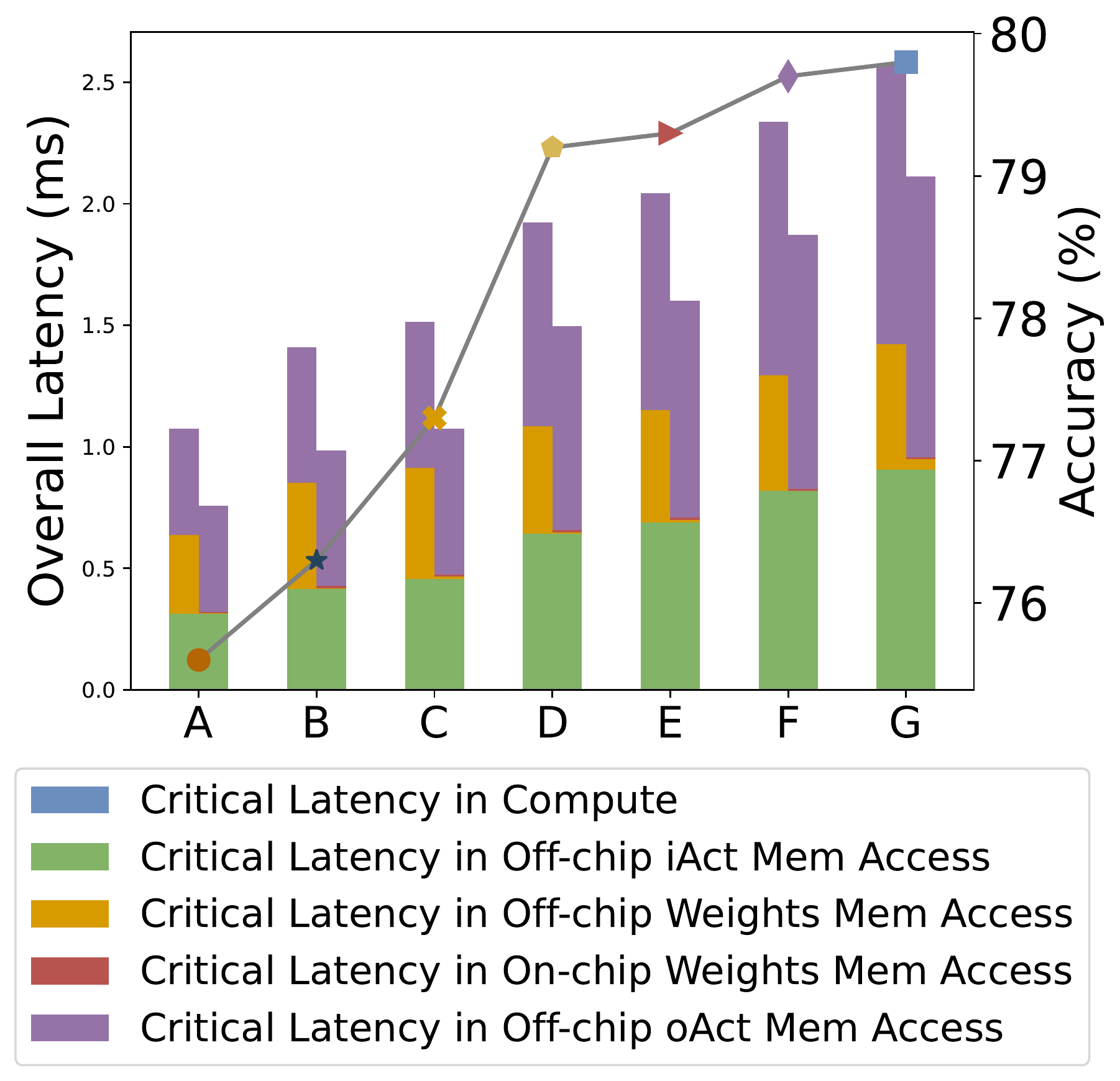}}} 
    \vspace{-4mm}
    \caption{Potential latency reduction with SGS (two bar per \subgraph, left: w/o PB; Right: w PB)}
    \vspace{-0.6cm}
    \label{fig:latency_accuracy_analysis_models}
\end{figure}

To understand the benefits of SGS, we perform roofline analysis as shown in \figref{fig:latency_accuracy_analysis_models} and \figref{fig:roofline_analysis_mem2comConvert}, where roofline represents the normal roofline curve. And SGS-roofline virtually improves the overall off-chip bandwidth by saving off-chip data access, leading to an improved roofline curve shown by SGS roofline. The experiments are performed with a system with 19.2 GB/s off-chip memory bandwidth and 1.296 Tflops throughput running at 100 MHz~\cite{AI_Accel_survey}.

The latency breakdown results in \figref{fig:latency_accuracy_analysis_models} shows that SGS can potentially remove the off-weights access latency from the critical path, such that the individual latency of serving a stream of queries from pareto-frontiers could be reduced by [6\%, 23.6\%] for MobV3 and [5.7\%, 7.92\%] for ResNet50. 

Such latency reduction essentially comes from the model boundedness shifting. The SGS pushes models towards compute-bound, which increases the utilization of the available compute resources for higher throughput and reduces latency and energy consumption. The shifting is illustrated by blue dots being pushed toward the red dots in \figref{fig:roofline_analysis_mem2comConvert}.

\begin{figure}[t]
    \centering
    \begin{subfigure}{0.22\textwidth}
        \centering
        \includegraphics[width=1.07\linewidth]{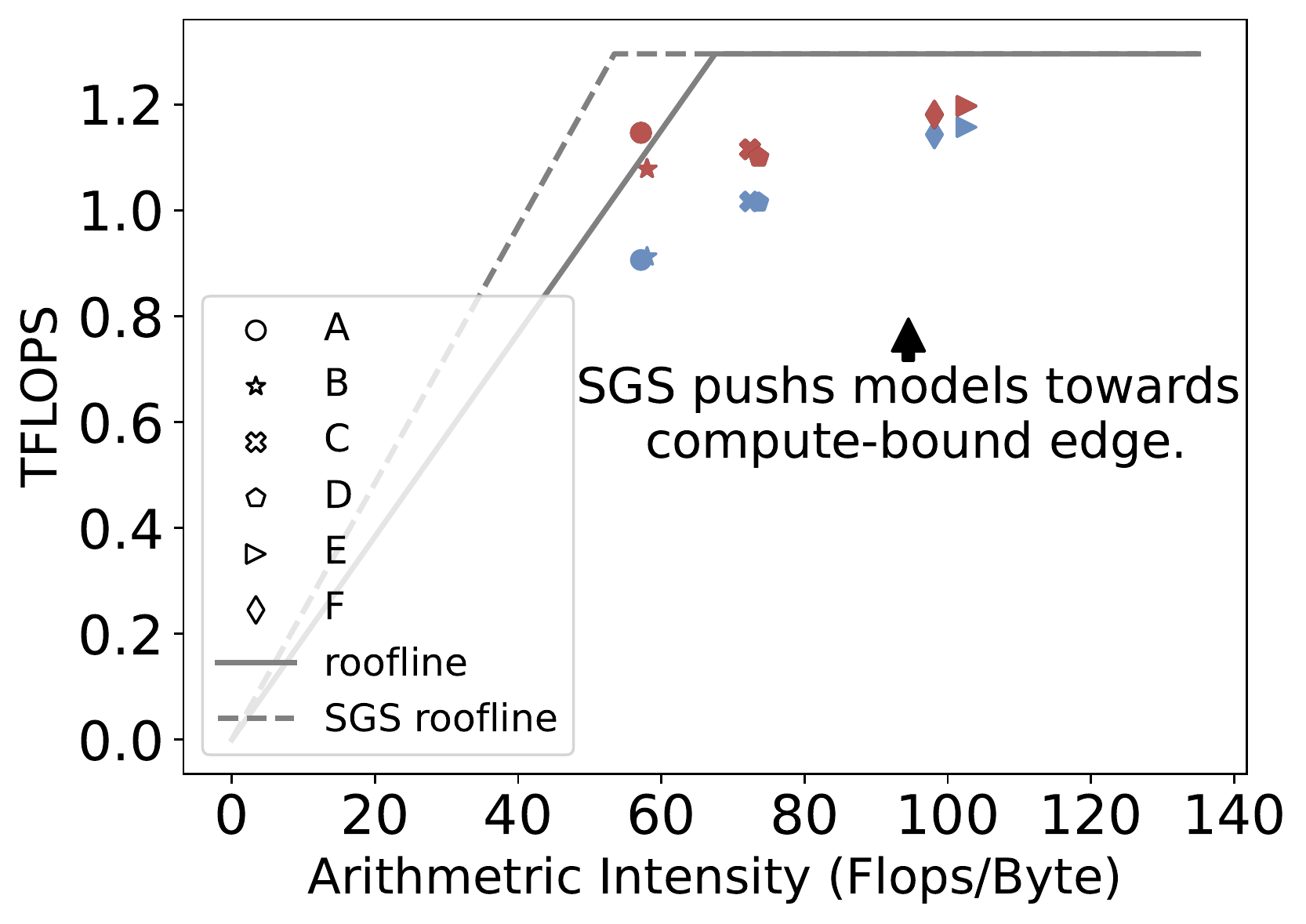}
        \caption{ResNet50.}
        \label{fig:roofline_resnet50}
    \end{subfigure}
    \begin{subfigure}{0.25\textwidth}
        \centering
        \includegraphics[width=0.95\linewidth]{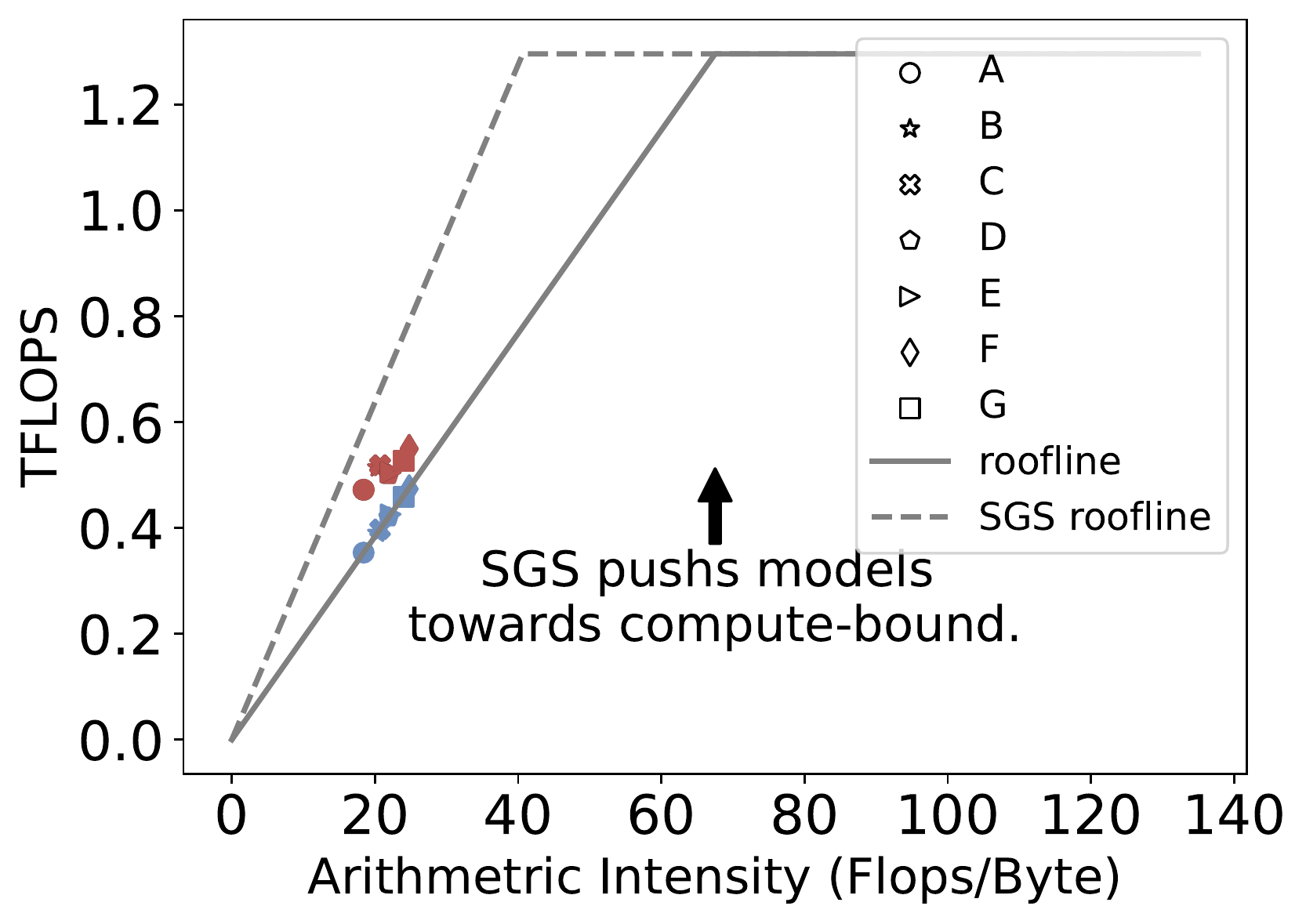}
        \caption{MobV3.}
        \label{fig:roofline_mbv3}
    \end{subfigure}
    \vspace{-4mm}
    \caption{SGS pushes memory-bound to compute-bound layers.}  
    \label{fig:roofline_analysis_mem2comConvert}
\end{figure}
\subsection{\systemhw Configuration Impact}

In this subsection, we explore the impact of three main factors (i.e., bandwidth, throughput, and PB size) of \systemhw on the overall end-to-end serving latency. 

\subsubsection{Bandwidth - Buffers Arrangement}
Different types of data require different bandwidths. A unified buffer for all different data types demands the controller to handle  potentially all-to-all connections between buffers and all compute units. While the design of the splitting buffer only needs a direct connection between a buffer and compute units, which saves the complexity of both the datapath and the controller. 
The buffer is a 2D array and its size equals  $width \times height$. The width refers to the bandwidth a buffer could supply every cycle.
The bandwidth demand of different buffers is shown in \tabref{tab:buffer_sizes}, which is determined by both workloads and hardware specifications.

\begin{table}[t]
    \centering
    \scriptsize
    \caption{Bandwidth requirement of on-chip buffers}
    \begin{tabular}{ccc}
        \hline
        Buffer & Minimal Bandwidth Requirement \\
        \hline
        DB     &  $LCM($ max off-chip $BW$, DPE Array demanded on-chip $BW$)                 \\
        SB     &  $LCM($ max off-chip $BW$, $C_P \times R \times S \times$ iActs DataWidth)  \\
        LB     &  DPE Array demanded on-chip $BW$  \\
        OB     &  $ K_P \times $ oAct DataWidth    \\
        PB     &  $LCM($ max off-chip $BW$, DPE Array demanded on-chip $BW$)     \\
        \hline
    \end{tabular}\\
    \label{tab:buffer_sizes}
    \noindent Note: $BW =$ bandwidth, $LCM(x_1,x_2)$: Least Common Multiple of $x_1$ and $x_2$.
    \vspace{-6mm}
\end{table}

\subsubsection{PB Size - Sizes of Buffers}

 All buffers compete on the same total storage budget so that a balance of them is preferred to achieve good performance.
The addition of a persistent buffer also introduces a new factor of common weights reuse, leading to 
a trade-off between inter-layer data reuse and intra-layer data reuse.

\subsubsection{Throughput - Parallelism of the Compute Array}
The parallelism of the 2D DPE Array is also a controllable knob. Within the same computation engine budget, a change in parallelism indicates a change in throughput, yielding different performances on different workloads. For example, the parallelism of $16$ and $32$ in K and C dimensions deliver a peak throughput of $512$ data per clock cycle. Therefore, we use this throughput as the factor to abstract parallelism.

\subsubsection{Design Space Exploration}

\begin{figure}[t]
    \centering
     \subfloat[ResNet50. \label{fig:ResNet50_DSE}]{{\includegraphics[width=0.495\columnwidth]{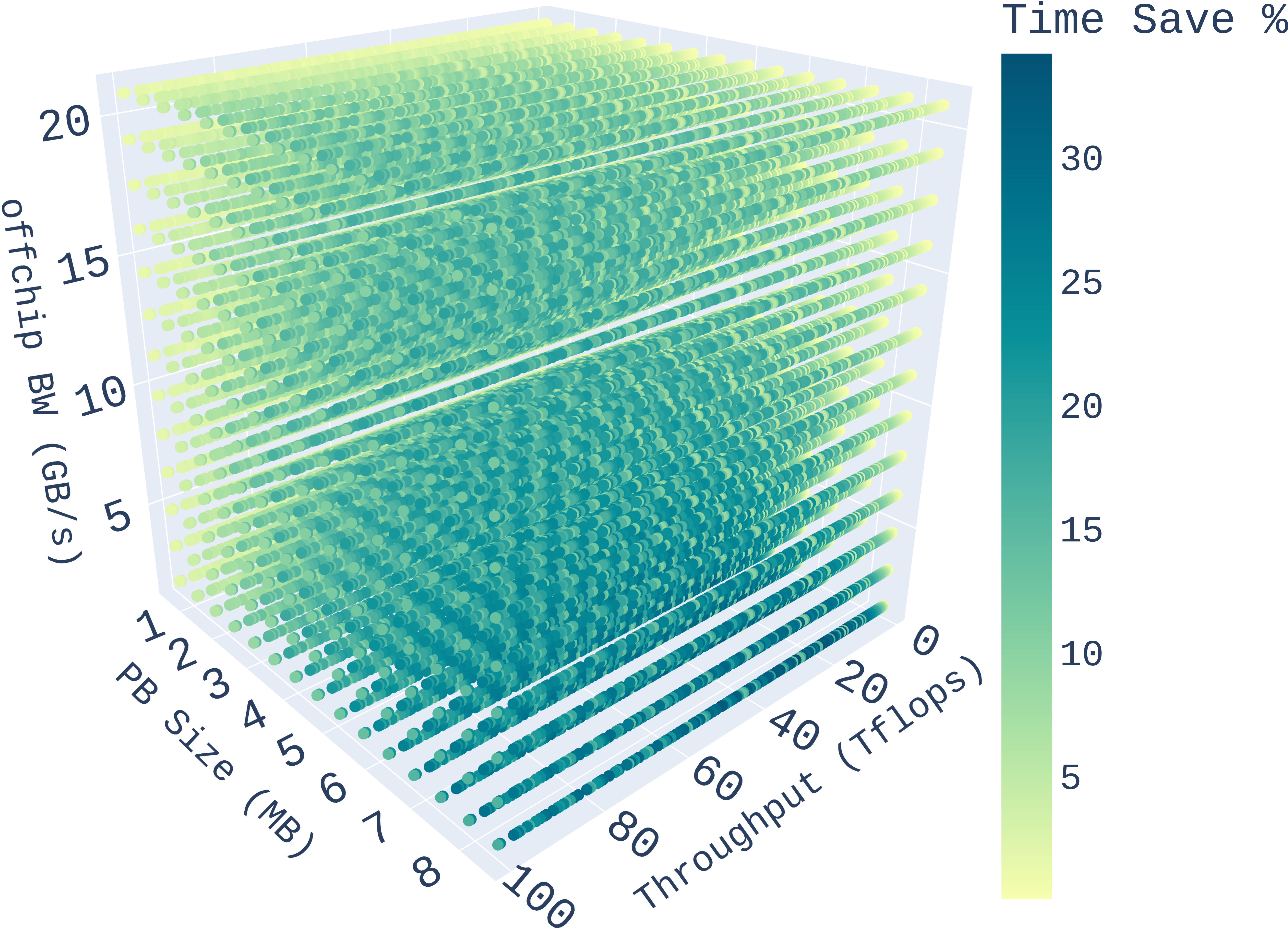}}}
    \vspace{-2mm}
    \subfloat[MobV3. \label{fig:MBv3_DSE}]{{\includegraphics[width=0.485\columnwidth]{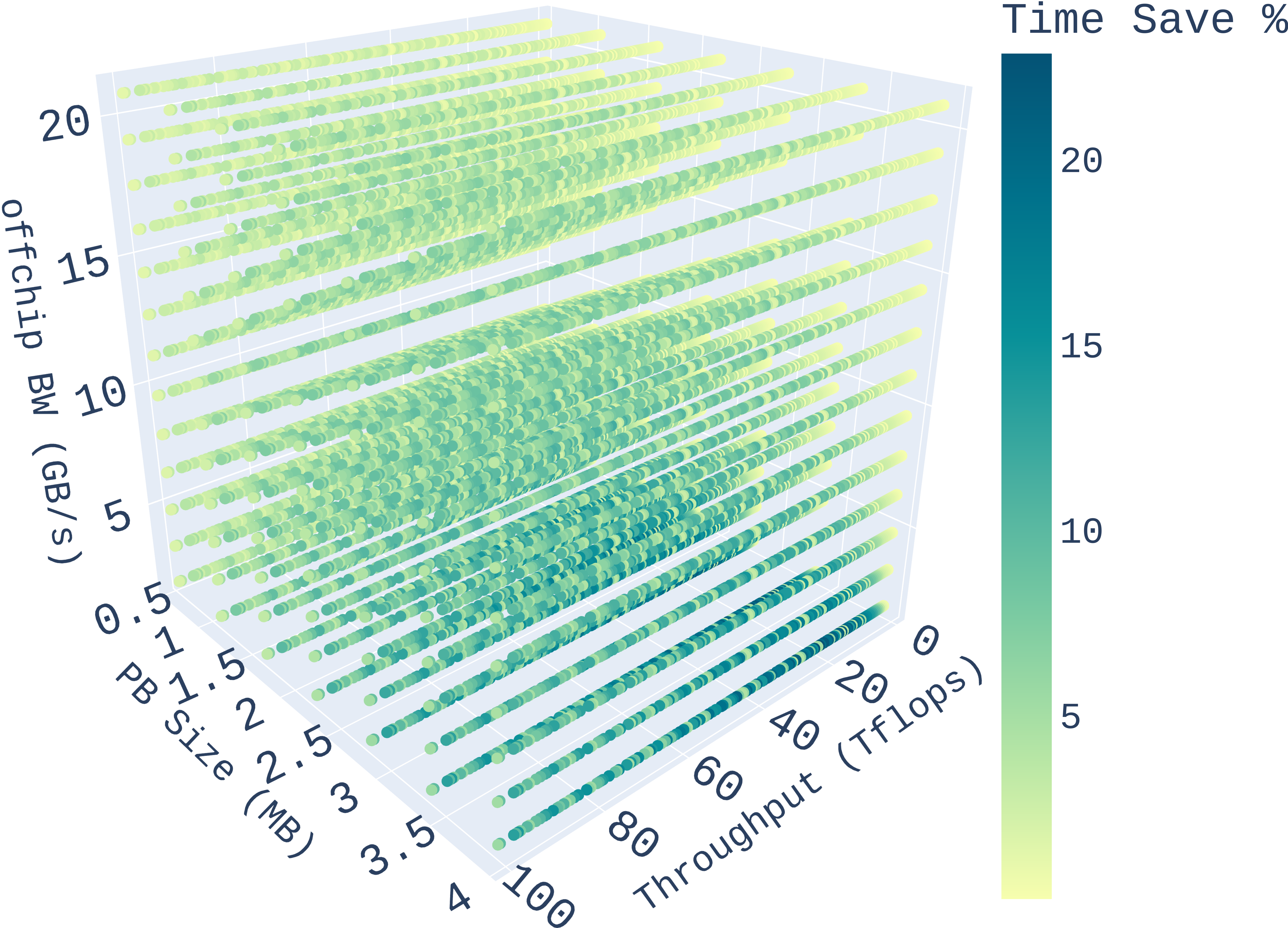}}} 
    \vspace{-2mm}
    \caption{Latency reduction (Time Save in legend) improvement exploration on \systemhw using Analytic Model.} 
    \label{fig:DSE}
\end{figure}

As~\figref{fig:DSE} shows with larger PB sizes, more on-chip computation, and less off-chip bandwidth, the latency is improved. However, for MobV3, due to the smaller size, having depth-wise conv layers, and less reuse, the amount of improvement is lesser for MobV3 compared with the ResNet50.

\subsection{\systemhw Evaluation}
In this subsection, we evaluate how \systemhw will impact the latency and energy reduction.
We evaluate different scales of \systemhw on two real FPGAs with different budgets running the 3x3 convolution layers of ResNet50. The \systemhw on Alveo U50 has off-chip bandwidth of 14.4 GB/s, PB size of 1.69 MB, and throughput of 0.9216 TFlops running at 100 MHz.

\subsubsection{Resources Allocation among Buffers}
\label{sec:buffer}
The resource utilization of \Accel w/ PB and w/o PB under optimal configurations on both Xilinx ZCU104 and Alveo U50 are shown in \tabref{tab:resource_zcu104} with a breakdown on-chip storage allocation shown in \tabref{tab:buffer}.
Both \Accel w/ PB and \Accel w/o PB use the same amount of overall on-chip storage for a fair comparison.
\vspace{-1mm}
\begin{table*}[t]
\centering
\vspace{-4mm}
\caption{{Resources comparison of \systemhw with DPU}}\label{tab:resource_zcu104}
\scriptsize 
\begin{tabular}{cccccccc}\hline 
               & \makecell{\Accel \\ w/o PB} & \makecell{\Accel \\ w/ PB}  & \makecell{Xilinx DPU \\ DPUCZDX8G} & \makecell{\Accel \\ w/o PB} & \makecell{\Accel \\ w/ PB}  \\\hline
Device         &      ZCU104      &  ZCU104  &    ZCU104  & Alveo U50      & Alveo U50   \\
LUT            &     61180 (26.6\%)    & 64307 (27.9\%)   &    41640 (18.1\%) & 231668 (26.63\%)  & 244969 (28.16\%) \\
Register       &    107216 (23.3\%)    & 117724 (25.5\%)  &    69180 (15\%)   & 435071 (24.96\%)  & 445602 (25.56\%) \\
BRAM           &     192.5 (61.7\%)    & 198.5 (63.6\%)   &      0            & 452.5 (33.67\%)   & 452.5 (33.67\%) \\
URAM           &      48 (50\%)      &  96 (100\%)  &     60 (62.5\%)         & 48 (7.5\%)        & 96 (15\%)\\
DSP            &     1507  (87.2\%)    & 1459 (87.2\%)    &     438 (25.35\%)   & 4739 (79.78\%)    & 4740 (79.79\%) \\
PeakOps/cycle  &     2592      & 2592   &    2304         &    9216           &  9216              \\
GFlops (100MHz) &     259.2     & 259.2  &     230.4      &    921.6          &  921.6             \\
\hline
\end{tabular}
\end{table*}

\vspace{-2mm}


\begin{table}[ht!]
\centering
\vspace{-3mm}
\caption{Buffer configurations of \Accel (ZCU104 board)}
\label{tab:buffer}
\scriptsize
\begin{tabular}{cccccc}
\hline
& \multicolumn{2}{c}{\Accel w/o PB} & \multicolumn{2}{c}{\Accel w/ PB} \\
& BRAM (KB) & URAM (KB) & BRAM (KB) & URAM (KB) \\
\hline
DB-Ping & 0 & 1152 & 0 & 576 \\
DB-Pong & 0 & 1152 & 0 & 576 \\
SB & 8 & 1152 & 8 & 576 \\
LB & 54 & 0 & 54 & 0 \\
OB & 327 & 0 & 327 & 0 \\
ZSB & 8 & 0 & 8 & 0 \\
PB & 0 & 0 & 0 & 1728 \\
Overall & 397 & 3456 & 397 & 3456 \\
\hline
\end{tabular}
\vspace{-4ex}
\end{table}

\subsubsection{Latency Evaluation}
\label{sec:self_evaluation}

The real-board latency and energy consumption results are shown in~\figref{fig:latency_resnet50} with resources shown in \tabref{tab:resource_zcu104}. On ZCU104, compared with CPU, \Accel w/o PB achieves $1.81X\sim 3.04X$ speedup and \Accel w/ PB achieves $1.87X\sim 3.17X$ for different \subnets. While on Alveo U50, compared with CPU, \Accel w/o PB achieves $1.43X\sim 2.54X$ speedup and \Accel w/ PB achieves $1.57X\sim 2.61X$ for different \subnets. 
\figref{fig:latency_resnet50} also shows that the scale-up design on Alveo U50 performs worse than the small-scale design on ZCU104 under small \subnets because of higher off-chip DRAM competition in data center cluster hosting Alveo U50 than simple embedded ZCU104. Thus, off-chip data access dominates latency in Alveo U50, resulting in the slow down for small \subnets.

\subsubsection{Energy Evaluation}
Energy in data movement has been proved to dominate the entire power consumption of neural network accelerator~\cite{dram_power} and thus we estimate the overall energy through profiling the off-chip DRAM data access for all different platforms shown in Figure~\ref{fig:energy_resnet50}. 

We estimate the off-chip energy by profiling the DRAM data access and compute it as $NumberAccess \times EnergyPerAccess$.
 With the proposed \subgraphreuse, we could save $[14\%, 52.6\%]$ off-chip data access energy saving for ResNet50 and $[43.6\%, 78.7\%]$ for MobV3 compared to \systemhw w/o PB. 

    
\begin{figure}[t]
    \centering
    \begin{subfigure}{0.235\textwidth}
        \centering
        \includegraphics[width=\linewidth]{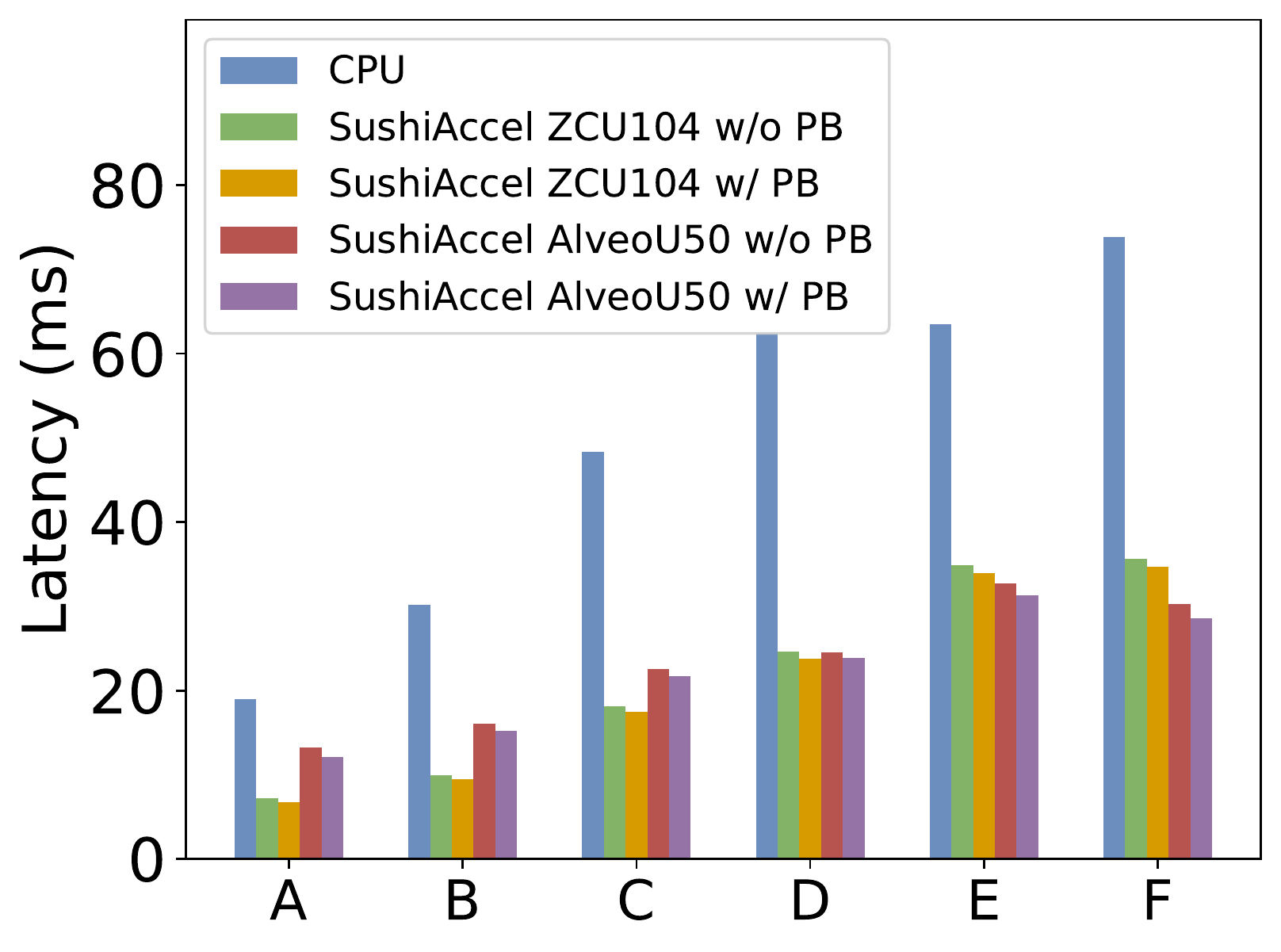}
        \caption{Latency comparison}
        \label{fig:latency_resnet50}
    \end{subfigure}
    \begin{subfigure}{0.24\textwidth}
        \centering
        \includegraphics[width=\linewidth]{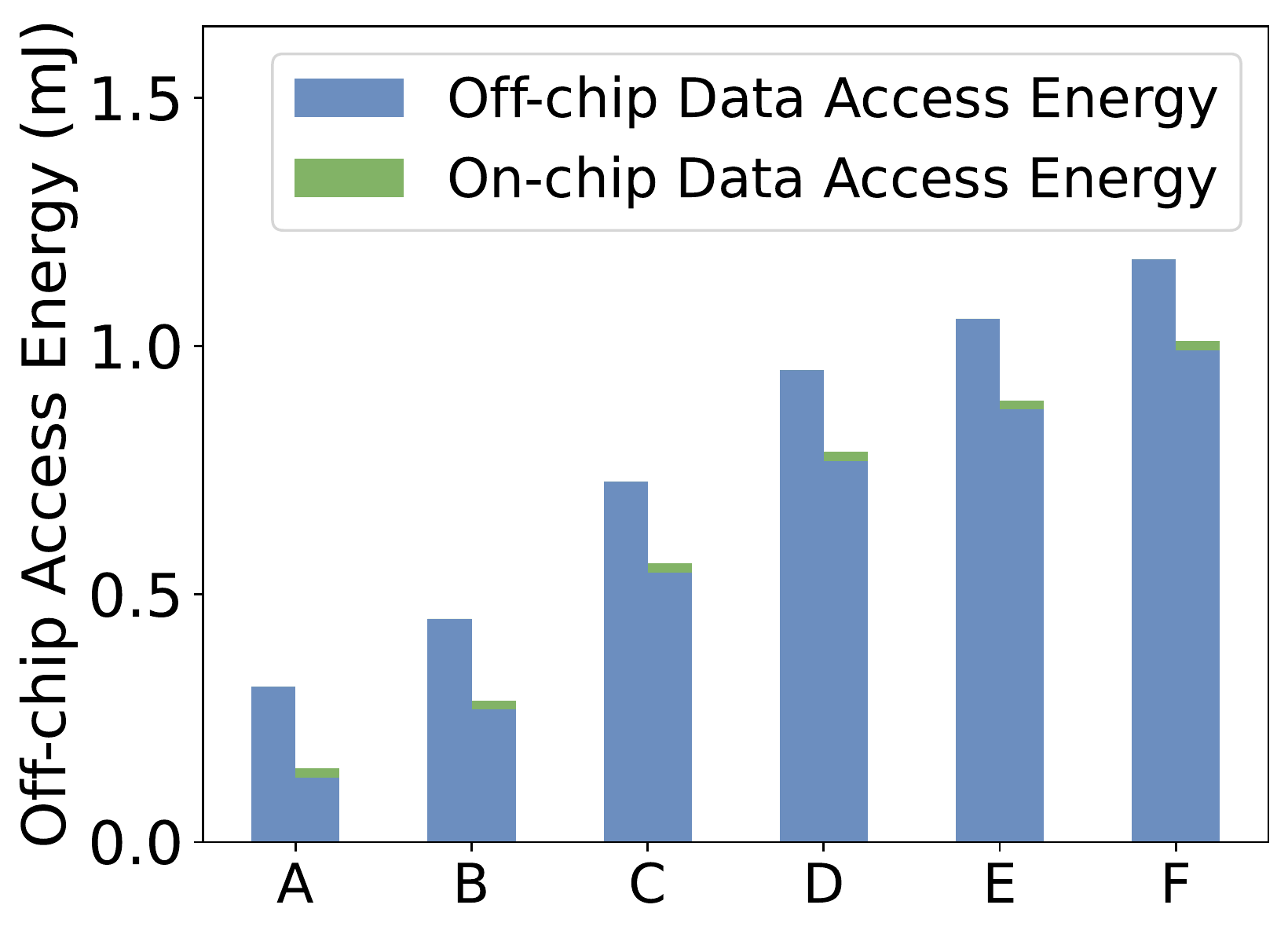}
        \caption{Energy comparison}
        \label{fig:energy_resnet50}
    \end{subfigure}
    \vspace{-6mm}
    \caption{Real board latency and energy reduction for ResNet50. (left and right bars in (b) are \systemhw w/o PB and w/ PB)}
    \label{fig:energy_analysis_models}
\end{figure}

\vspace{-2mm}
\subsection{Comparing with DPU}
\label{sec:xilinx}
We compared \Accel against Xilinx DPU using real layer-wise end-to-end inference latency of min-\subnet on ZCU104 as shown in \figref{fig:Latency_Comparison}. We consider convolution layers with $3\times 3$ kernel sizes. \Accel w/o PB achieved 0.5$\sim$1.95$\times$ faster execution time than Xilinx DPU ($25.1\%$ GeoMean speedup). 
This quantitative comparison lends credence to the proposal of adding a Persistent Buffer (PB) to a state-of-the-art ML accelerator design.
 
There are also seldom cases when \Accel performs worse than Xilinx DPU, because \Accel takes less parallelism in height (X) and width (Y) dimensions (\figref{fig:terminology}), leading to higher latency under workload with higher X and Y values.

\begin{figure*}[tb]
\begin{center}
    \vspace{-4mm}
    \includegraphics[scale=0.4]{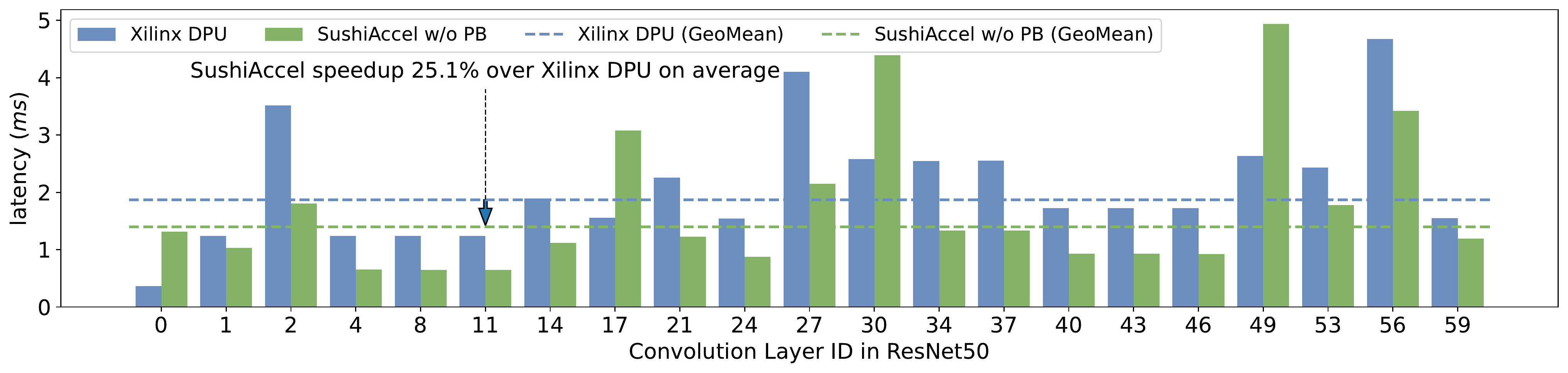}
    \vspace{-4mm}
    \caption{The latency comparison between \Accel w/o PB and Xilinx DPU for ResNet50.}
    \label{fig:Latency_Comparison}
\end{center}
\vspace{-6mm}
\end{figure*}

\vspace{-2ex}
\subsection{\systemsw Functional Evaluation}
In this section, we evaluate the performance of \systemsw for both ResNet50 and MobV3.

\begin{figure*}[t]
    \centering
    \vspace{-2mm}
    \subfloat[Latency of ResNet50. \label{fig:lesserLatResNe50}]{{\includegraphics[width=0.51\columnwidth]{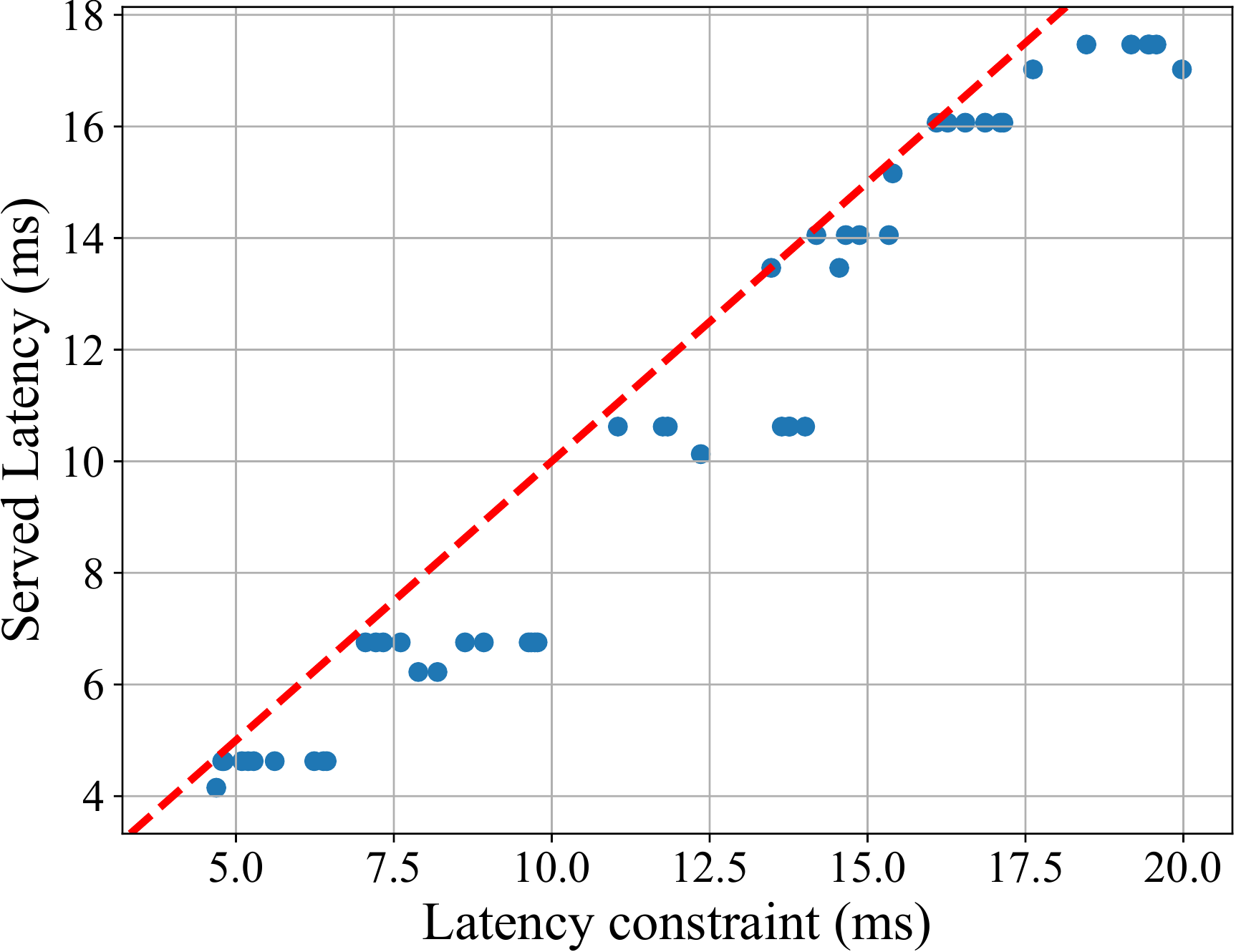}}} 
    \subfloat[Accuracy of ResNet50. \label{fig:betterAcResNe50}]{{\includegraphics[width=0.53\columnwidth]{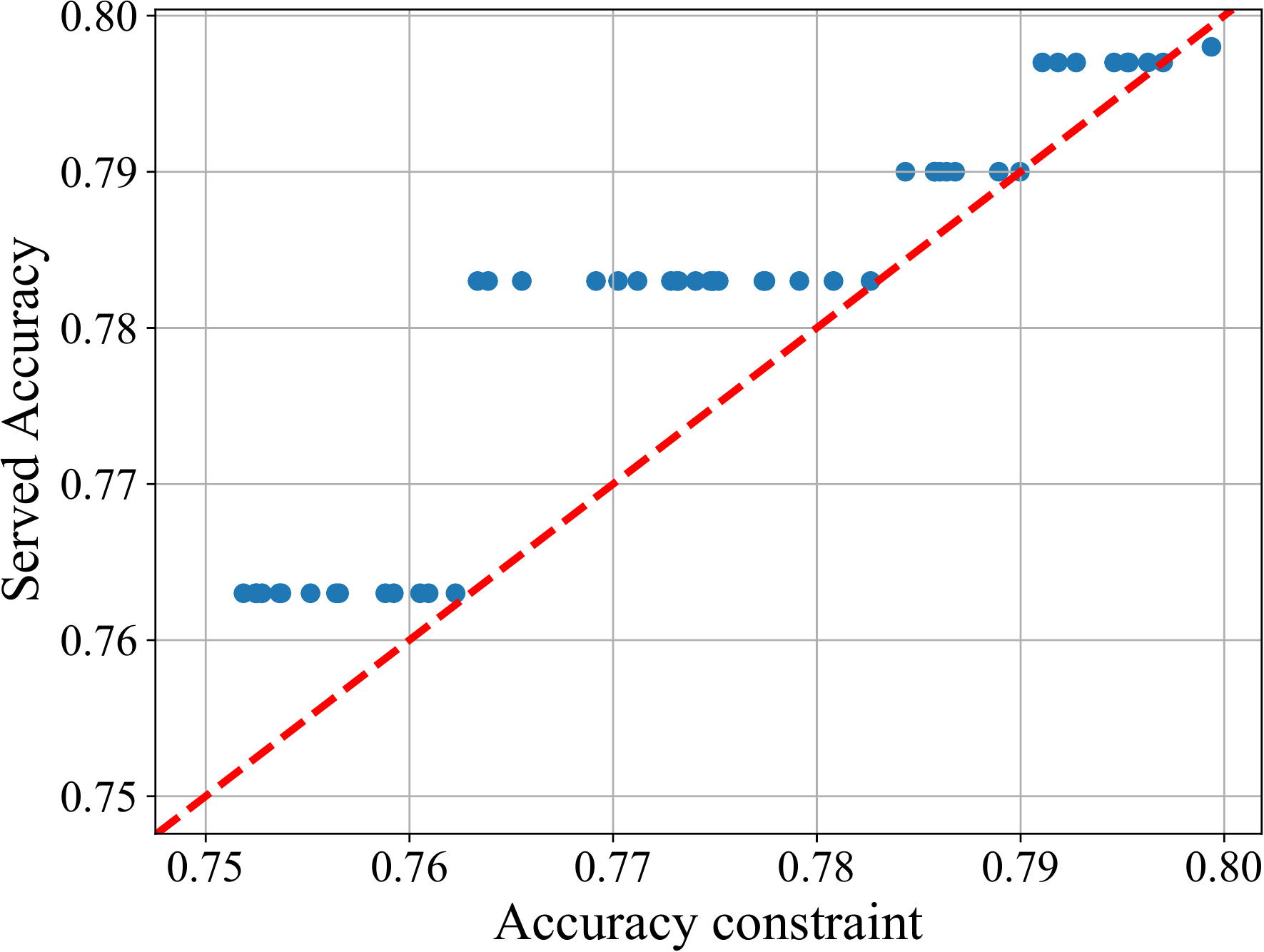}}}
    \subfloat[Latency of MobV3. \label{fig:lesserLatMobv3}]{{\includegraphics[width=0.51\columnwidth]{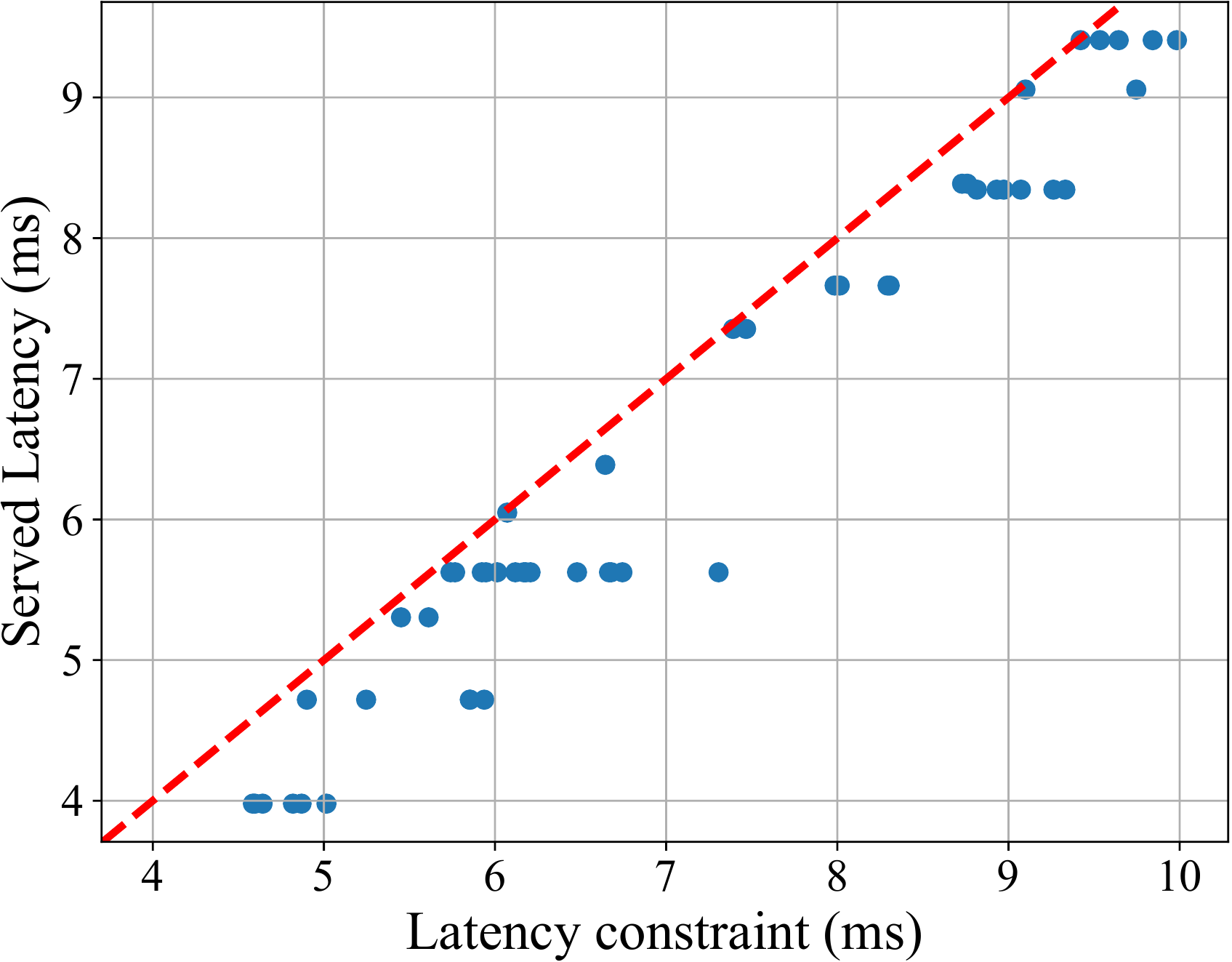}}} 
    \subfloat[Accuracy of MobV3. \label{fig:betterAcMobv3}]{{\includegraphics[width=0.53\columnwidth]{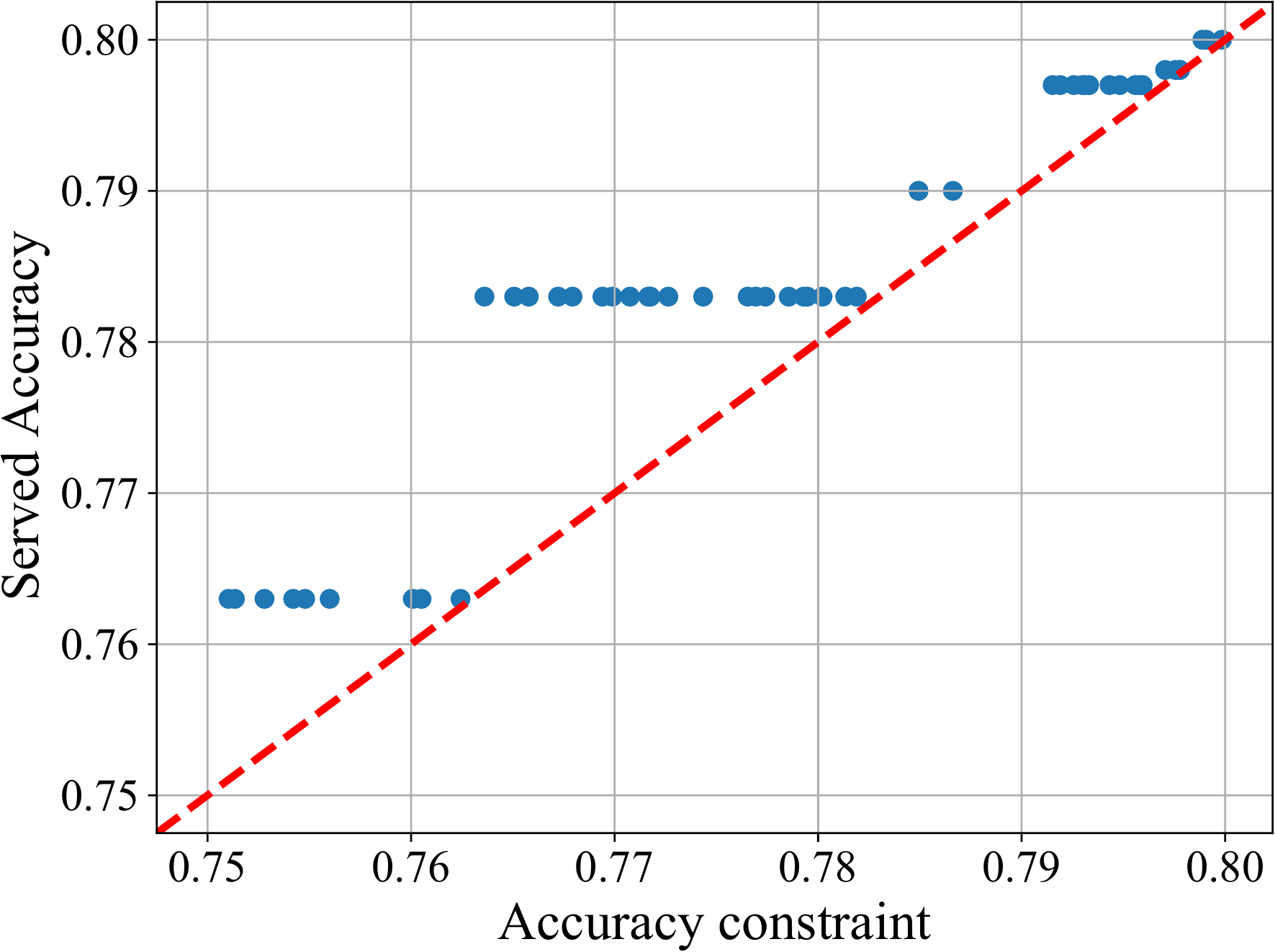}}}
    \vspace{-3mm}
    \caption{Serve strictly better accuracy and lesser latency for ResNet50 and MobV3 using \system.}
    \label{fig:latency_accuracy_troff}
    \vspace{-4mm}
\end{figure*}

\figref{fig:latency_accuracy_troff} shows that the \systemsw is able to serve queries with strictly lesser latency and/or better accuracy where blue dots represent served queries by employing \systemsw. In ~\figref{fig:lesserLatResNe50} and ~\figref{fig:lesserLatMobv3}, blue dots are almost always below the line $y=x$ manifesting that the \systemsw can serve strictly lesser latency if the latency is a hard constraint that needs to be satisfied. Similarly, all blue dots above the line $y=x$ in ~\figref{fig:betterAcResNe50} and \figref{fig:betterAcMobv3} show that the \systemsw can serve strictly better accuracy if accuracy is a hard constraint that needs to be met. 

\vspace{-2ex}
\subsection{End-to-End \system Evaluation}

In this section, we compare the latency-accuracy tradeoff results among \system w/o PB, \system w/ PB (state-unaware caching), and \system. The blue dots in \figref{fig:comparison-latency-accuracy} illustrate how \system serves random queries\footnote{Due to the overlap, only limited points in the figures are visible}. 

{
For ResNet50 in all cases, \system w/o scheduler
consistently {outperforms} {No-\system}. For random queries, {\system} is also able to decrease the latency by 21\% on average given the same accuracy compared to not having \system. }

{In the case of MobV3, due to its small size, a relatively larger fraction of a \subnet fits in PB, resulting in a higher cache-hit ratio {(Appendix~\ref{sec:app:cachehit})}. 
\system offers better accuracy-latency tradeoff than \system w/o scheduler, with the exception of only a few points. 
In the case of MobV3, {\system} is also able to decrease the latency by 25\% on average given the same accuracy compared to not having \system.} 

Finally, \system increases the {{serving}} accuracy by up to 0.98\% for the same latency, which is significant for ML serving applications.

\begin{figure}[t]
    \centering
    \vspace{-2mm}
   \subfloat[ResNet50. \label{fig:SushiLatvsACcompResNet}]{{\includegraphics[width=0.5\linewidth]{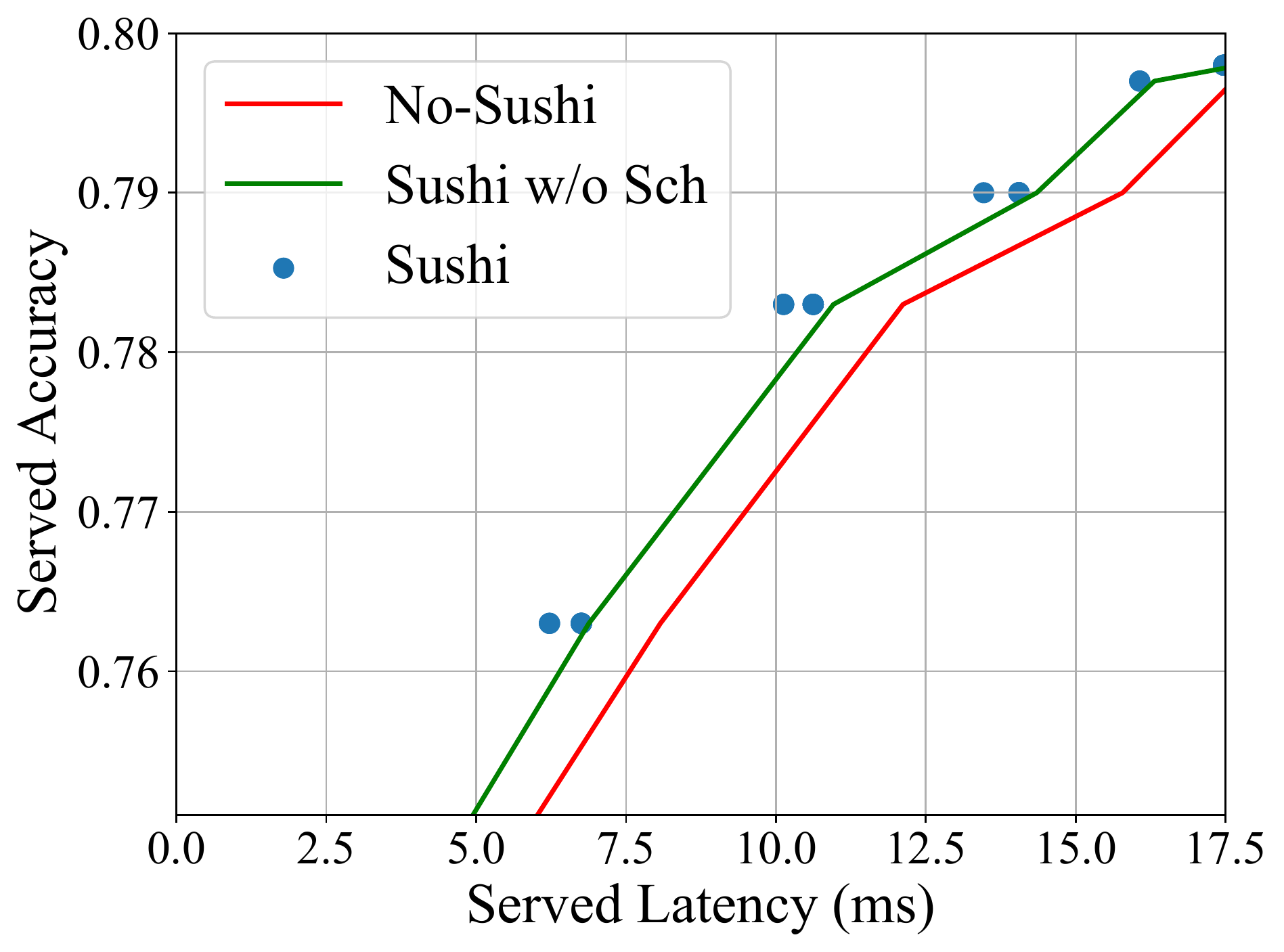}}}
    \subfloat[MobV3. \label{fig:SushicompRtMobV3}]{{\includegraphics[width=0.475\linewidth]{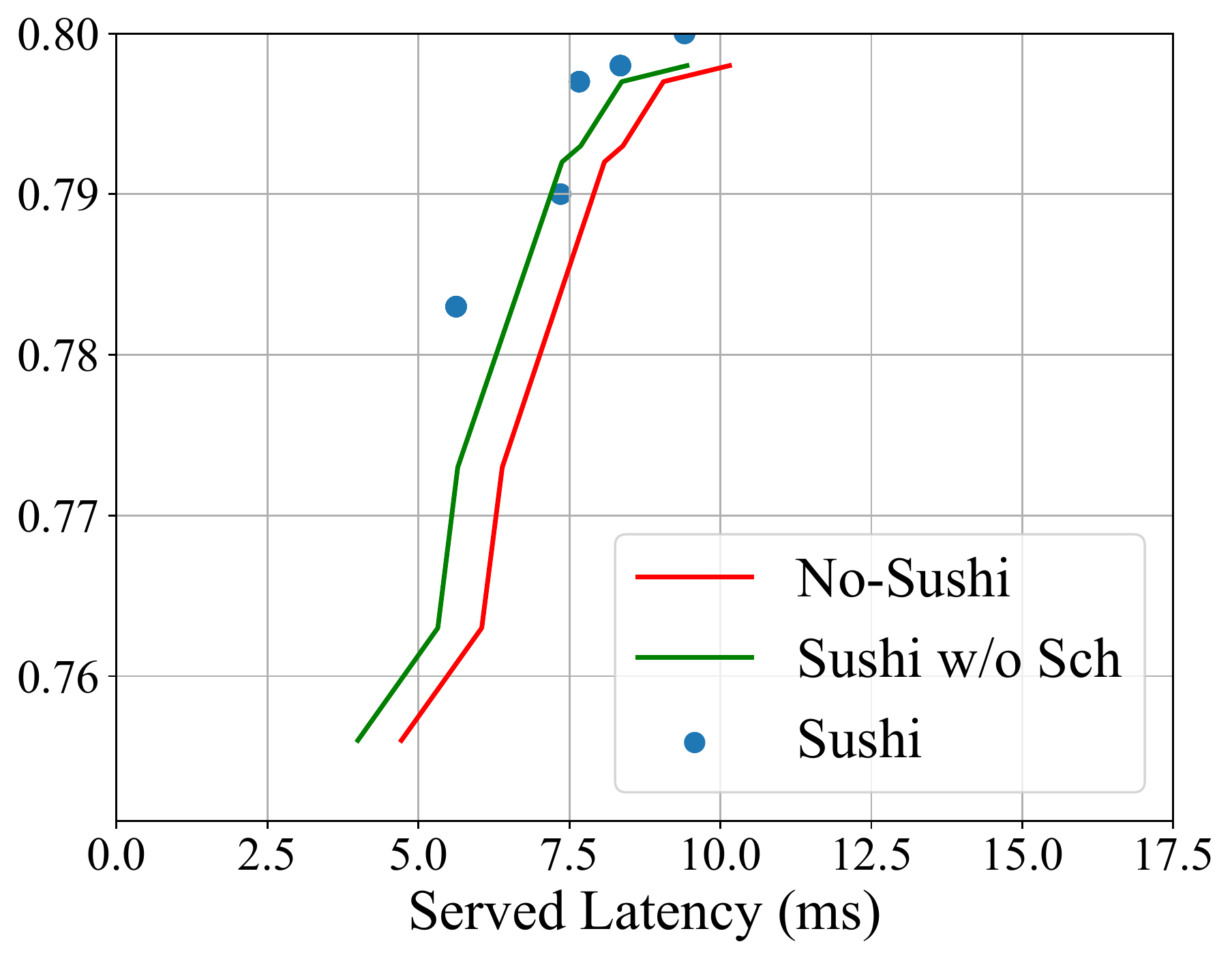}}} 
    \vspace{-4mm}
    \caption{Comparing delivering latency-vs-accuracy of No-\system and \system w/o scheduler and baselines for ResNet50 and MobV3.}
    \vspace{-5mm}
    \label{fig:comparison-latency-accuracy}
    \vspace{-1.5ex}
\end{figure}

\section{Related Work}
\label{sec:related}

\begin{table}[t]
    \centering
    \scriptsize
    \vspace{-5mm}
    \caption{Reuse comparison (prior works v.s. \system).}
    \resizebox{\columnwidth}{!}{
    \begin{tabular}{ccccc}
        \hline
            Work &  \makecell{iActs Reuse \\ \figref{fig:sliding_win_reuse} \& \ref{fig:kernel_reuse}} &  \makecell{oAct Reuse \\ Partial Sum} & \makecell{Weights Reuse \\ iAct Tiling} &   \makecell{\subgraph \\Reuse} \\
            \hline
            MAERI~\cite{kwon2018maeri} &\cmark  & \xmark & \cmark &  temporal \xmark \\
            NVDLA~\cite{nvdla}  & \xmark & \cmark & \cmark &  temporal \xmark \\
            Eyeriss~\cite{chen2016eyeriss}    &\cmark & \xmark & \cmark & temporal \xmark \\
            Xilinx DPU~\cite{xilinx_dpu}  &\cmark  & \cmark & \cmark & temporal \xmark  \\
            \system  &\cmark &\cmark &\cmark & \makecell{spatial  \cmark\\ temporal \cmark}  \\
        \hline
    \end{tabular}}
    \vspace{-5mm}
    \label{tab:cmp_other_accel}
\end{table}

Various accelerator designs such as Maeri~\cite{kwon2018maeri}, Eyeriss~\cite{ chen2018eyeriss}, NVDLA~\cite{nvdla}, and DPU~\cite{xilinx_dpu} support different types of reuse~\figref{fig:workload_data_reuse}. A comparison of them is shown in \tabref{tab:cmp_other_accel}. However, all of these works achieve intra-model cross-layer reuse in contrast to the cross-query reuse we propose with \systemhw.

Clipper~\cite{crankshaw2017clipper} serves single model queries without exposing a latency/accuracy tradeoff.
Inferline ~\cite{crankshaw2018inferline} serves multiple models but in a pipeline, there's no latency/accuracy tradeoff per model.
INFaaS~\cite{romero2021infaas} provides a query-time latency/accuracy tradeoff mechanism and policy but suffers from expensive model switching mechanisms. This also translates into a policy that minimizes model switching as a result.
The vertically integrated inference serving stack provided by \system naturally plugs into existing inference serving frameworks, enabling
 agile navigation of the latency/accuracy tradeoff at query time. 

\section{Conclusion}
\label{sec:con}
\system{} is a vertically integrated hardware-software inference serving stack that takes advantage
of the temporal locality induced by serving inference queries on the same weight-shared supernetwork structure.
To the best of our knowledge, the concept of SubGraph Stationary (SGS) optimization across queries is novel.
We demonstrate that to achieve the best temporal locality benefit, the proposed hardware implementation \systemhw{} must work in tandem with the software scheduler \systemsw
to control what SubNets to serve for each query and how to update the accelerator state. 
We further ensure generalizability of \systemsw by abstracting the effect of hardware state on the latency (and energy) of served SubNets with a black box SubGraph latency table. 
This decouples \systemsw from any accelerator implementation, while maintaining its state-awareness implicitly. 
\system can be naturally integrated in state-of-the-art ML inference serving frameworks and enables better latency/accuracy tradeoffs for a stream of queries with latency/accuracy constraints. 
{For a stream of queries, our results show \shepherd{0.98\%} improvement in the served accuracy, and up to 25\% latency reduction.}

\section{Acknowledgment}
\label{sec:ack}

{This material is based upon work partially supported by the National Science Foundation under Grant Number CCF-2029004. Additional support was provided by a sponsored research award by Cisco Research. 
We would like to further acknowledge the insightful comments of the review panel as well as the skillful guidance of our shepherd,  Dr. Qijing Jenny Huang, which greatly contributed to the quality of this paper.
We thank the anonymous reviewers of MLSys, and the SAIL Research Group members for valuable feedback and the stimulating intellectual environment they provide.
We also thank Taekyung Heo from Synergy lab for his feedback on the initial version of the paper.
{\bf Disclaimer: } Any opinions, findings, and conclusions or recommendations expressed in this material are those of the authors and do not necessarily reflect the views of the National Science Foundation.

\bibliographystyle{mlsys2023}
\bibliography{main}

\newpage
\newpage
\appendix
\section{Appendix - Ablation Studies}
\label{sec:appendix}

\vspace{-3mm}
\subsection{Temporal Analysis of Subgraph Caching}
\vspace{-2mm}
In this section, we explore the impact of a number of vectorized \subgraphs employed in the running average results as well as the size of $Latency-Table$ on the accuracy-latency results. Making cache update decisions after each query improves both latency and accuracy results (\figref{fig:Temporal_analysis_ResNet50}), but is prohibitively expensive as the new \subgraph must be fetched from off-chip memory.
 
For ResNet50, increasing the number of queries to two, the results worsen. Increasing the number of queries to 4 and 8 yields better results. Eventually, there's a point when the performance starts to get worse (e.g., at 10+ queries) as the benefit of temporal locality will be reduced. So there's a tradeoff between the staleness of query history over which the cached \subgraph is computed and the cost of updating cache frequently.

Following the same methodology for MobV3 (\ref{fig:temporal-analysis-MobV3}), we observe that averaging over 10 queries gives us the best tradeoff, leading to better accuracy-latency results.

\vspace{-4mm}
\subsection{Impact of $Latency-Table$ size}

\begin{table}[b]
\vspace{-0.40in}
\centering
\caption{Average latency improvement with respect to the size of  $Latency-Table$ normalized to \system w/o scheduler}
\label{tab:latTabsize}
\resizebox{0.8\columnwidth}{!}{
\begin{tabular}{cccccc}
\hline
 &10-cols& 40-cols&  80-cols& 100-cols &500-cols     \\ \hline
 ResNet50& 4\% & 7\%& 8\% &  9\%&  9\%    \\ 
 MobV3& 1\%& 1\%& 1\% & 1\% & 1\%    \\ \hline
\end{tabular}
}
\vspace{-0.33in}
\end{table}

The results in \tabref{tab:latTabsize} show the average latency improvement by increasing the size of $Latency-Table$ compared with {\system w/o scheduler}. As the results for ResNet50 show, increasing the size of the table improves performance, but is quickly saturated.
This is consistent with the important property of \systemsw table (rapid lookups on the critical query path). 

For MobV3, we see almost no improvement in latency with increased table size, which shows that if the PB is large enough to hold a large portion of the \subnet (and, with other on-chip buffers---the whole \subnet), the small table size can capture most of the required information by the scheduler. 
Thus, for smaller models, we keep the horizontal size of $Latency-Table$ minimal.

\vspace{-4mm}
\subsection{Lookup Latency}
\vspace{-1mm}

We used the lookup table as a fast-search data structure. For the largest model, (ResNet-50) the latency in microseconds is shown in table \tabref{tab:latencylookup}. These results show that the lookup table time is less than $\frac{1}{1000}$ of the inference time and, thus, doesn't significantly interfere with the query's critical path.

\begin{table}[t]
\centering
\caption{Look up time (us)}
\label{tab:latencylookup}
\resizebox{0.8\columnwidth}{!}{
\begin{tabular}{cccccc}
\hline
 &100-cols& 200-cols&  500-cols& 1000-cols &2000-cols     \\ \hline
 ResNet50& 2 & 4 & 6 & 10 &  17 \\   
 \hline
\end{tabular}
}
\vspace{-0.31in}
\end{table}

\vspace{-4mm}
\subsection{Cache Hit Ratio}
\vspace{-2mm}

\label{sec:app:cachehit}
\system leverages temporal locality across queries as the SubNets they induce share some weights that are common to these SubNets. The benefit of \subgraphreuse thus fundamentally is a function of the workload. For instance, if all queries used the exact same SubNet and we cache the largest SubGraph of that SubNet, then the probability of its reuse is 1.
To generalize this intuition, \system makes caching decisions based on the intersection of SubNets used by the last $Q$ queries.
Thus, we define the cache hit ratio as the fraction of the cached SubGraph that was ``hit'' or present in the SubNet served, because those weights don't need to be fetched.
 
For a given query trace, we log $(SN_t, G_t)$ series of tuples where $SN_t$ is the \subnet that the scheduler decided to serve at time $t$, and $G_t$ is a \subgraph cached in PB at time $t$. We find the overlap between $SN_t$ and $G_t$ using $\frac{\|(SN_t \cap G_t)\|_2} {\| (SN_t)\|_2}$ where $SN_t$ is already vectorized using $C$ and $K$. We average this over $t$ to get the average cache hit ratio. $||\cdot||_2$ is used as a proxy to calculate vector overlap. Thus defined, \system reaches a hit ratio of 66\% (78\%) for ResNet50 (MobV3). 
It is instructive that the cache hit ratio is higher for smaller models, as the intersection of common weights used by \subnets over a past window of $Q$ queries is a larger fraction of the served \subnet.

\begin{figure}[ht!]
    \centering
    \vspace{-2ex}
    \begin{subfigure}{0.23\textwidth}
        \centering  \includegraphics[width=\linewidth]{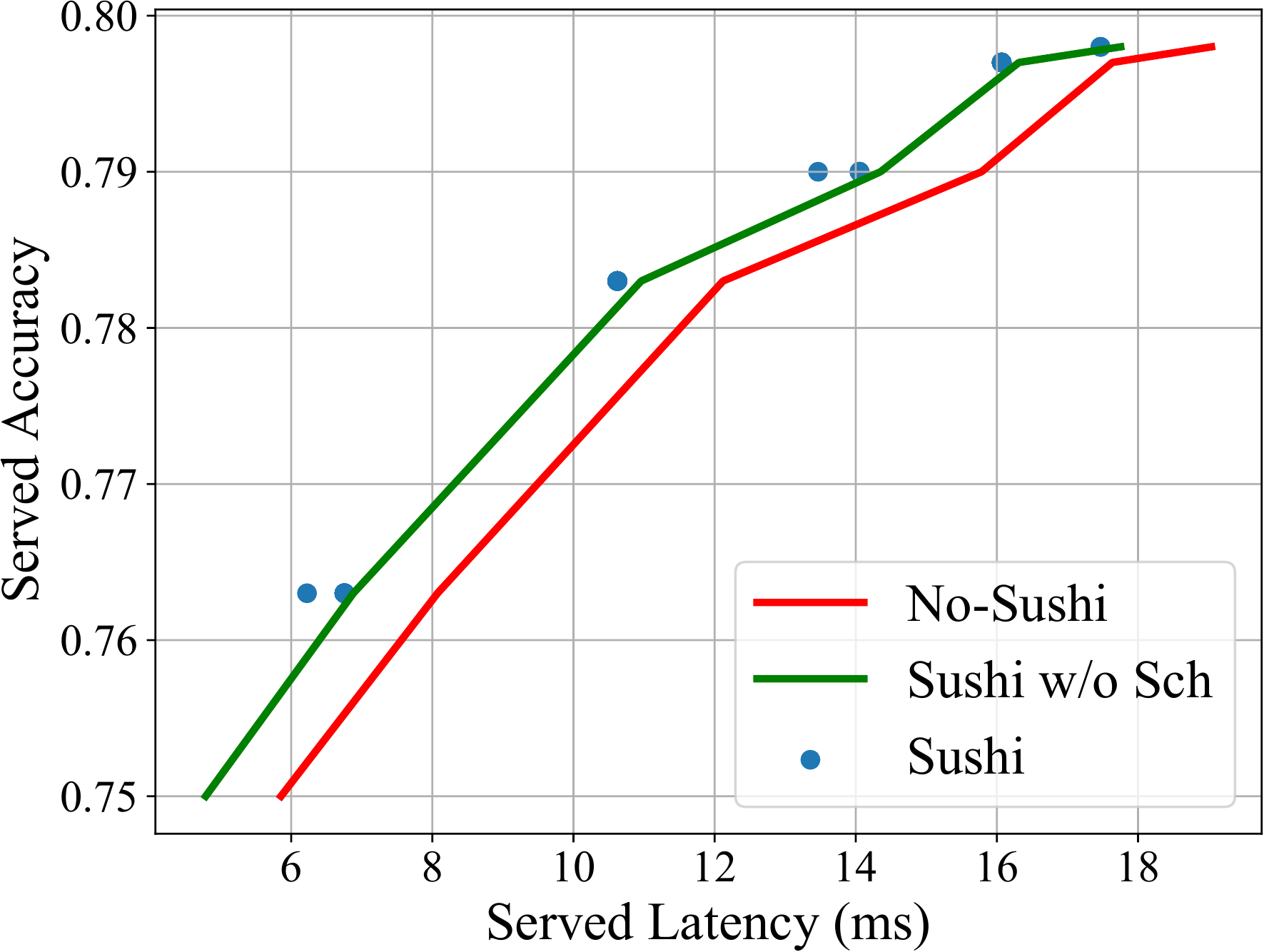}
        \vspace{-4ex}
        \caption{Single query}
        \label{fig:sgl_query_resnet50}
    \end{subfigure}
    \begin{subfigure}{0.23\textwidth}
        \centering
        \includegraphics[width=\linewidth]{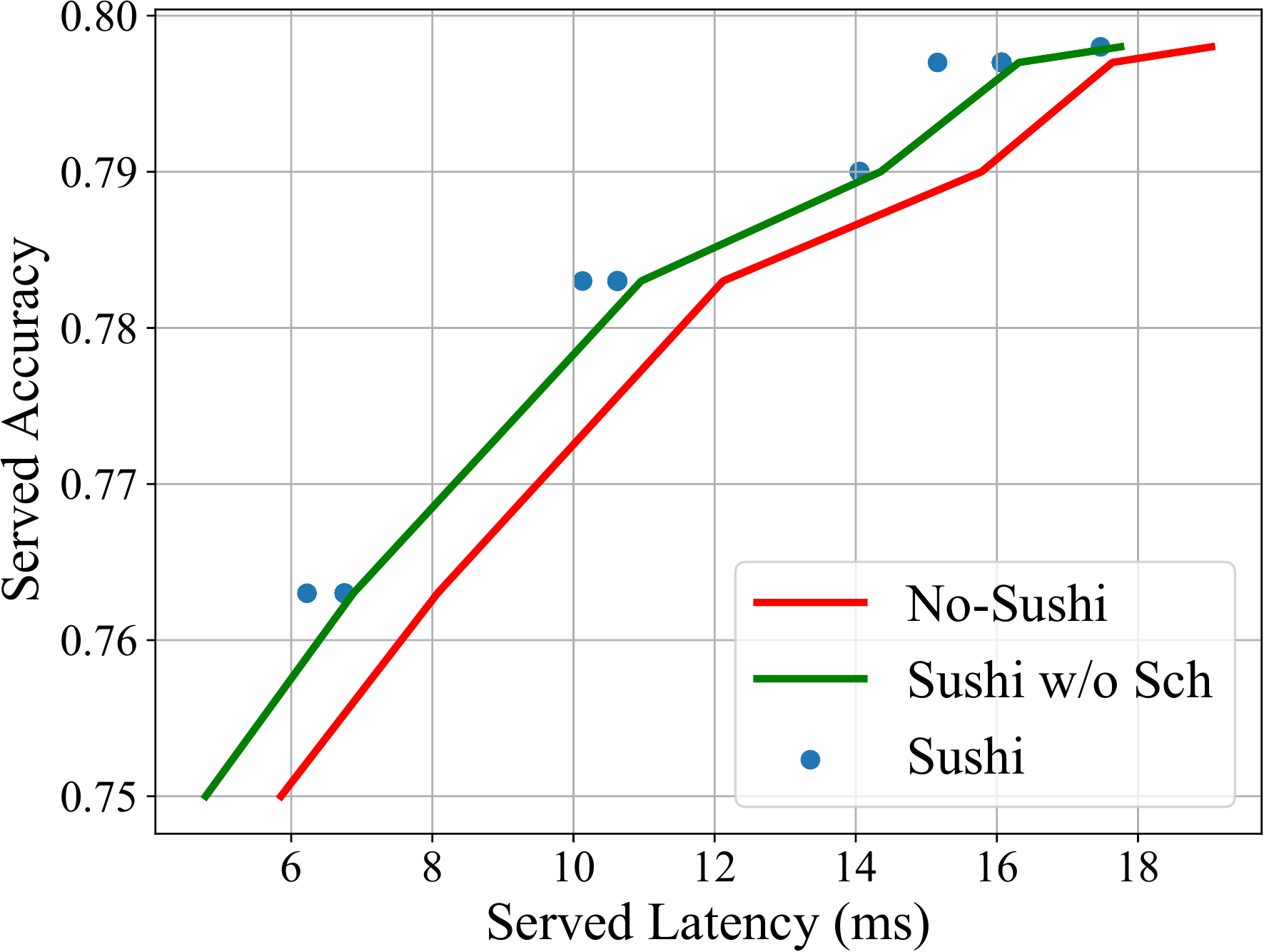}
        \vspace{-4ex}
    \caption{Averaging 2 queries} \label{fig:sgl_query_resnet50_avg2}
    \end{subfigure}
    \begin{subfigure}{0.23\textwidth}
        \centering       \includegraphics[width=\linewidth]{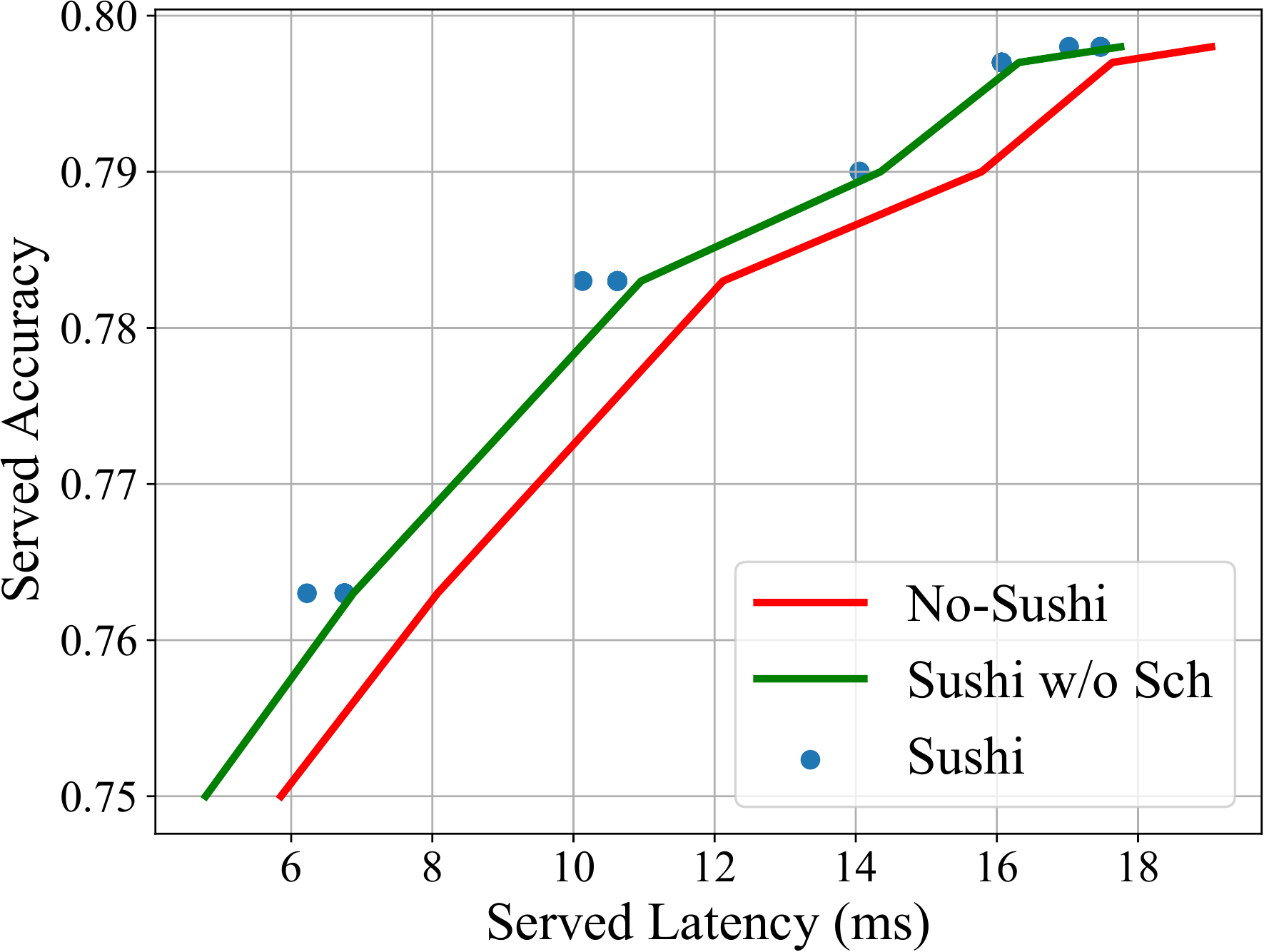}
        \vspace{-4ex}
    \caption{Averaging 4 queries}    \label{fig:sgl_query_resnet50_avg4}
    \end{subfigure}
    \begin{subfigure}{0.23\textwidth}
        \centering
        \includegraphics[width=\linewidth]{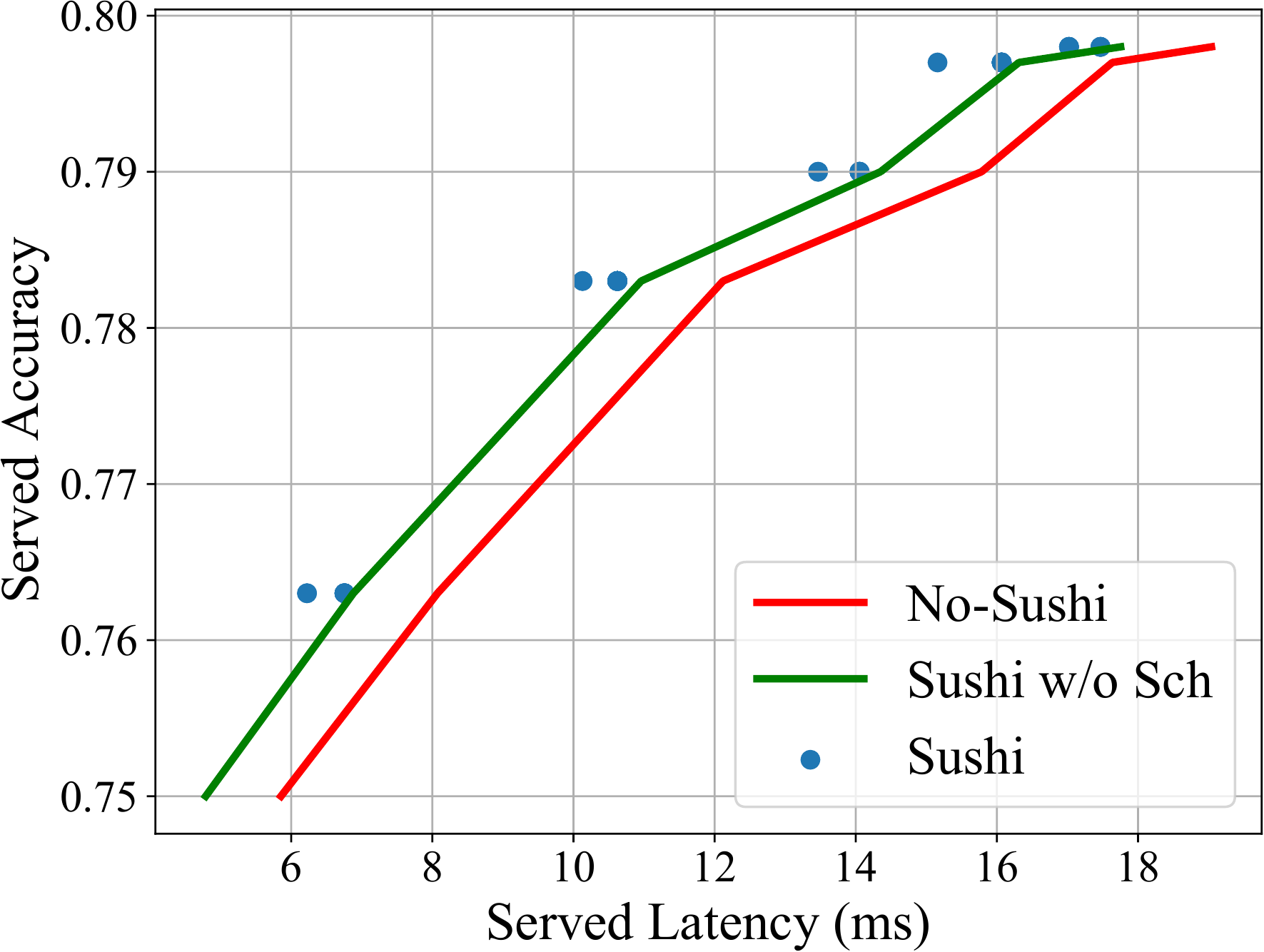}
        \vspace{-4ex}
        \caption{Averaging 10 queries}       \label{fig:sgl_query_resnet50_avg10}
    \end{subfigure}
    \vspace{-2ex}
    \caption{Temporal analysis of subgraph caching for ResNet50}  \label{fig:Temporal_analysis_ResNet50}
   \vspace{-2ex}
\end{figure}

 \vspace{-1ex}
\begin{figure}[ht!]
    \centering
    \begin{subfigure}{0.22\textwidth}
        \centering
        \includegraphics[width=1\linewidth]{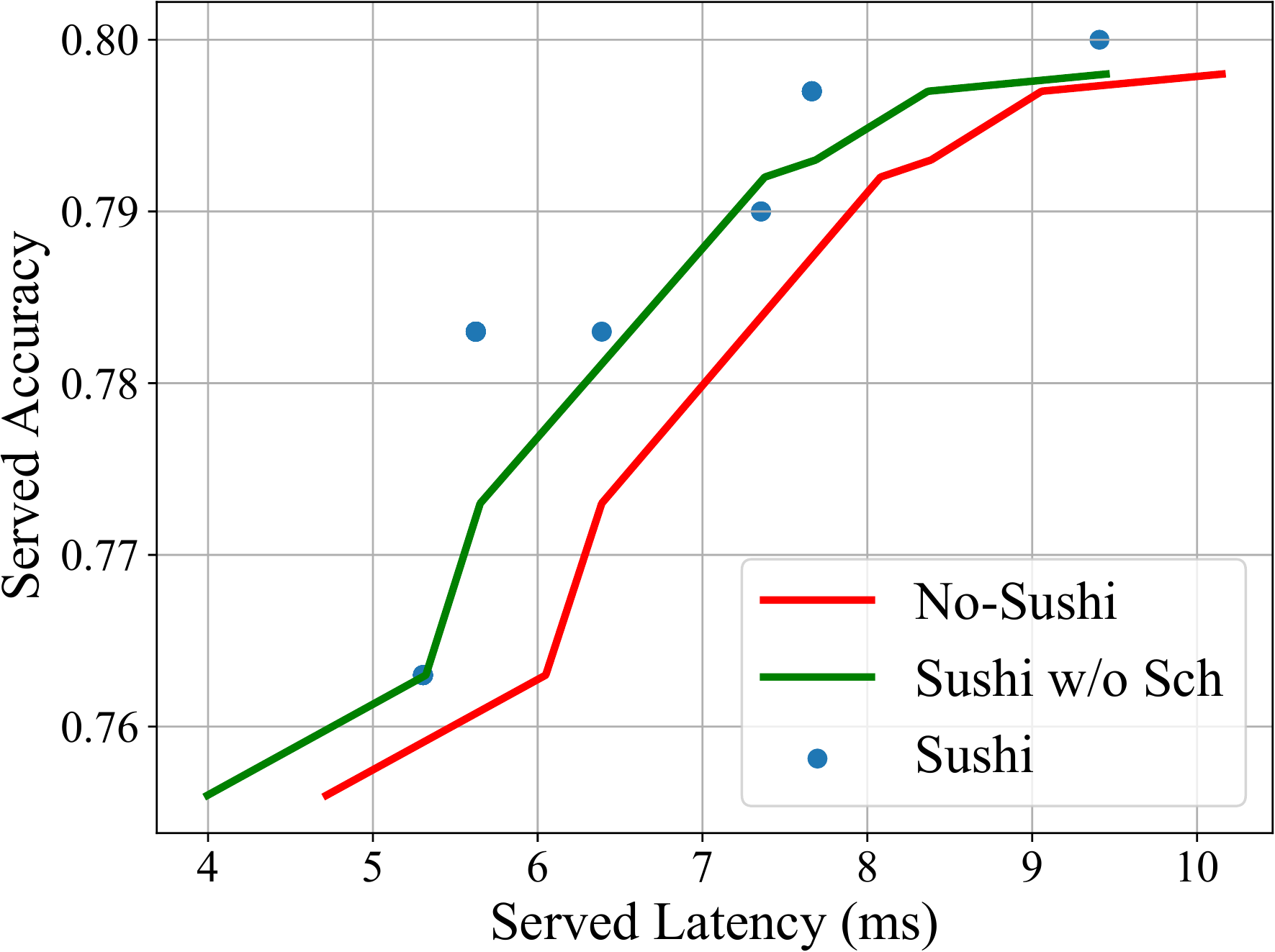}
         \vspace{-4ex}
        \caption{Single query}
        \label{fig:sgl_mobv3}
    \end{subfigure}
    \begin{subfigure}{0.22\textwidth}
        \centering
        \includegraphics[width=\linewidth]{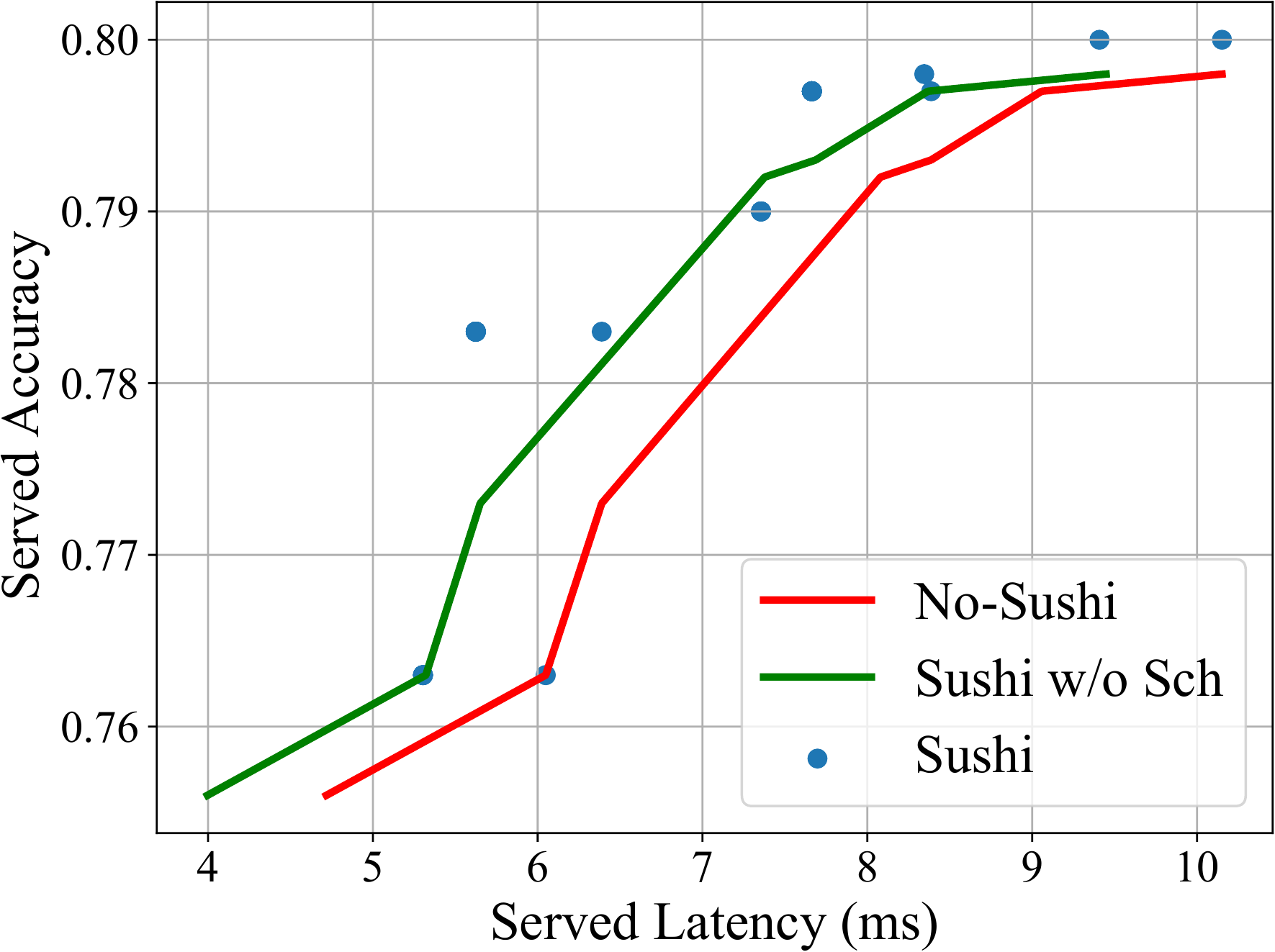}
         \vspace{-4ex}
        \caption{Averaging 4 queries}
        \label{fig:avg4_mobv3}
    \end{subfigure}
    \begin{subfigure}{0.22\textwidth}
        \centering
        \includegraphics[width=\linewidth]{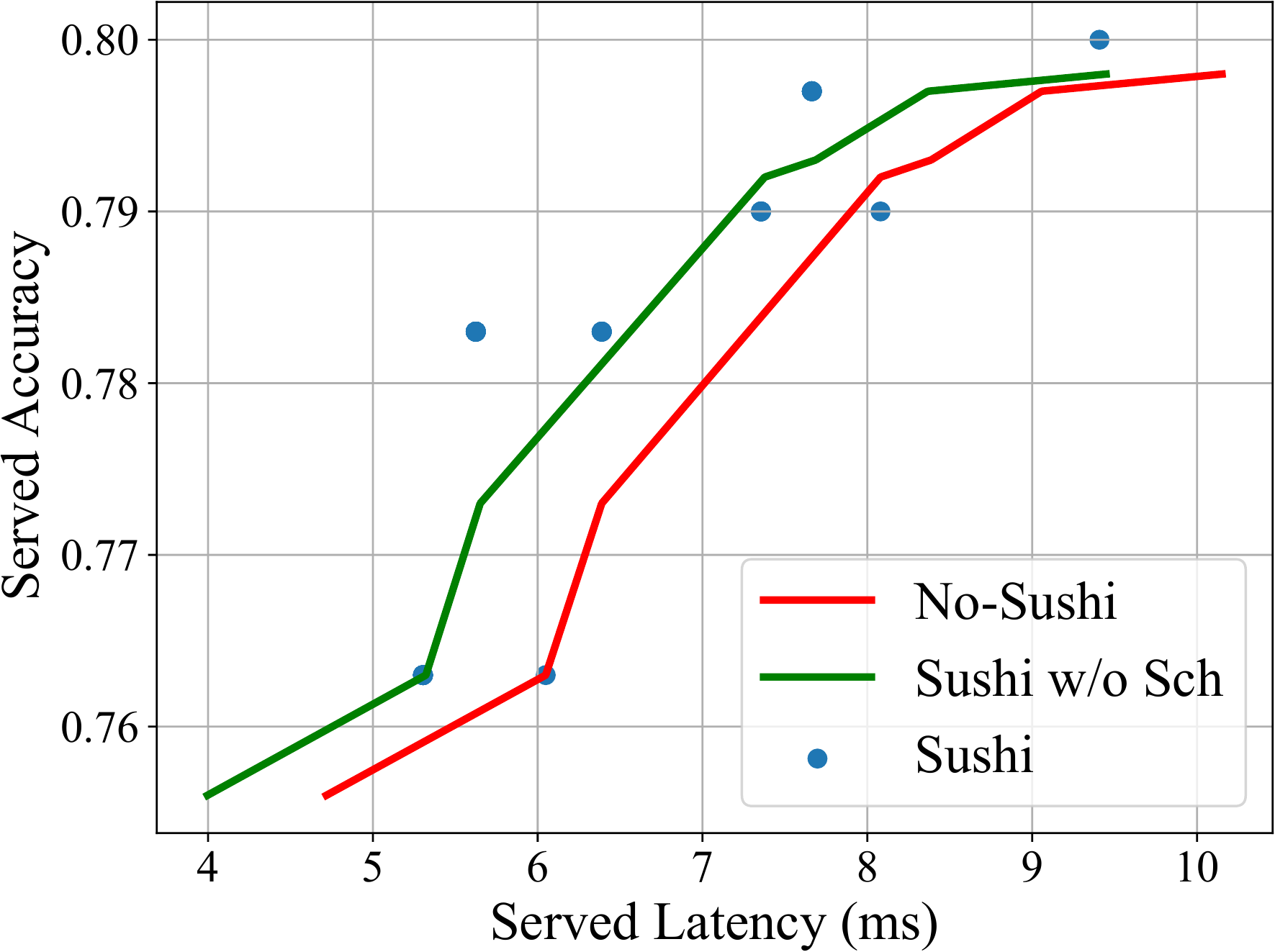}
         \vspace{-4ex}
        \caption{Averaging 8 queries}
        \label{fig:avg8_mobv3}
    \end{subfigure}
    \begin{subfigure}{0.22\textwidth}
        \centering
        \includegraphics[width=1\linewidth]{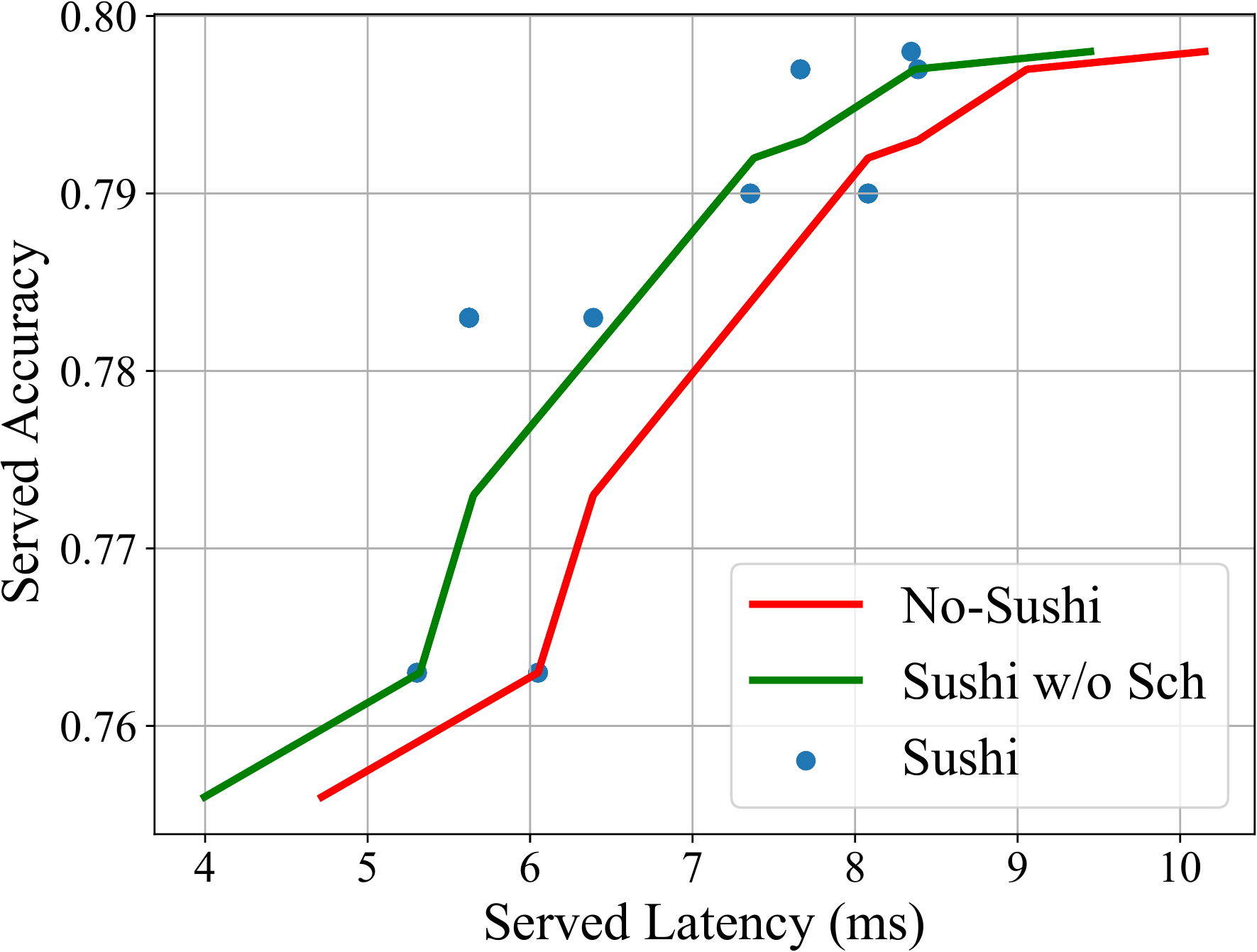}
         \vspace{-4ex}
        \caption{Averaging 15 queries}
        \label{fig:avg15_mobv3}
    \end{subfigure}
    \vspace{-4mm}
    \caption{Temporal analysis of subgraph caching for MobV3}
    \label{fig:temporal-analysis-MobV3}
    \vspace{-4mm}
\end{figure}

\end{document}